\documentclass[aps,article,twocolumn,longbibliography,citeautoscript,footinbib,superscriptaddress,prl]{revtex4-2}
\usepackage[T1]{fontenc}
\usepackage{feynmp}
\usepackage{comment}
\usepackage{subcaption}
\usepackage{amsmath}
\usepackage{amssymb}
\usepackage{amsthm}
\usepackage{amsfonts}
\usepackage{bm,bbm}
\usepackage{braket}
\usepackage{xcolor}
\usepackage{pifont}
\usepackage{slashed}
\newcommand{\be}{\begin{equation}}
\newcommand{\ee}{\end{equation}}
\usepackage[mathscr]{euscript}
\usepackage[shortlabels]{enumitem}
\usepackage{tikz-feynman}
\usepackage[colorlinks=true]{hyperref}
\hypersetup{
    bookmarks=true,         % show bookmarks bar?
    unicode=false,          % non-Latin characters 
    pdftoolbar=true,        % show Acrobat
    pdfmenubar=true,        % show Acrobat 
    pdffitwindow=false,     % window fit to page when opened
    pdfstartview={FitH},    % fits the width of the page to the window
    pdfsubject={},   % subject of the document
    pdfcreator={},   % creator of the document
    pdfproducer={}, % producer of the document
    pdfkeywords={} {} {}, % list of keywords
    pdfnewwindow=true,      % links in new window
    colorlinks=true,       % false: boxed links; true: colored links
    linkcolor=magenta, %red,          % color of internal links (change box color with linkbordercolor)
    citecolor=blue,        % color of links to bibliography
    filecolor=magenta,      % color of file links
    urlcolor=blue           % color of external links
} 

\usepackage{amsfonts}
\usepackage{upgreek}
\usepackage{slashed}
\usepackage{latexsym}

\newcommand{\beq}{\begin{equation}}
\newcommand{\eeq}{\end{equation}}
\def\bea{\begin{eqnarray}}
\def\eea{\end{eqnarray}}
\newcommand{\nn}{\nonumber \\}

\DeclareUnicodeCharacter{2212}{-}
\DeclareUnicodeCharacter{03B4}{$\delta$}

\renewcommand{\approx}{\simeq}

\definecolor{wrongultramarine}{rgb}{1,0.5,0}

\newcommand{\rd}{{\rm d}}
\newcommand{\dd}{{\rm d}}

\newcommand{\iw}{i\omega}
\newcommand{\om}{\omega}

\usepackage{xr}
\makeatletter
\newcommand*{\addFileDependency}[1]{
 \typeout{(#1)}
 \@addtofilelist{#1}
  \IfFileExists{#1}{}{\typeout{No file #1.}}
}
\makeatother

\newcommand*{\myexternaldocument}[1]{
    \externaldocument{#1}
    \addFileDependency{#1.tex}
    \addFileDependency{#1.aux}
}
\myexternaldocument{supp5}
\usepackage{pdfpages}

\makeatletter
\AtBeginDocument{\let\LS@rot\@undefined}
\makeatother

\begin{document}

\begin{flushright}
 \href{https://arxiv.org/abs/2406.07608}{arXiv:2406.07608} 
\end{flushright}

\title{Strange metal and superconductor in the\\ two-dimensional Yukawa-Sachdev-Ye-Kitaev model}

\author{Chenyuan Li}
\affiliation{Department of Physics, Harvard University, Cambridge MA 02138, USA}
\author{Davide~Valentinis}
\affiliation{Institut f\"ur Quantenmaterialien und Technologien, Karlsruher Institut
f\"ur Technologie, 76131 Karlsruhe, Germany}
\affiliation{Institut f\"ur Theorie der Kondensierten Materie, Karlsruher Institut
f\"ur Technologie, 76131 Karlsruhe, Germany}
\author{Aavishkar A. Patel}
\affiliation{Center for Computational Quantum Physics, Flatiron Institute, New York,
New York, 10010, USA}
\author{Haoyu~Guo}
\affiliation{Laboratory of Atomic and Solid State Physics, Cornell University,
142 Sciences Drive, Ithaca NY 14853-2501, USA}
\author{J\"org Schmalian}
\affiliation{Institut f\"ur Quantenmaterialien und Technologien, Karlsruher Institut
f\"ur Technologie, 76131 Karlsruhe, Germany}
\affiliation{Institut f\"ur Theorie der Kondensierten Materie, Karlsruher Institut
f\"ur Technologie, 76131 Karlsruhe, Germany}
\author{Subir Sachdev}
\affiliation{Department of Physics, Harvard University, Cambridge MA 02138, USA}
\author{Ilya Esterlis}
\affiliation{Department of Physics, University of Wisconsin-Madison, Madison, Wisconsin 53706, USA}

\date{\today
\\
\vspace{0.4in}}

%%%%%%%%%%%%%%%%%%%%%%%%%%%%%%%%%%%%%%%%%
\begin{abstract}
The two-dimensional Yukawa-Sachdev-Ye-Kitaev (2d-YSYK) model provides a universal theory of quantum phase transitions in metals in the presence of quenched random spatial fluctuations in the local position of the quantum critical point. It has a Fermi surface coupled to a scalar field by spatially random Yukawa interactions. 
We present full numerical solutions of a self-consistent disorder averaged analysis of the 2d-YSYK model in both the normal and superconducting states, obtaining electronic spectral functions, frequency-dependent conductivity, and superfluid stiffness. Our results reproduce key aspects of observations in the cuprates as analyzed by Michon {\it et al.} (Nat. Comm. {\bf 14}, 3033 (2023)). We also find a regime of increasing zero temperature superfluid stiffness with decreasing superconducting critical temperature, as is observed in bulk cuprates. 
  \end{abstract}
%%%%%%%%%%%%%%%%%%%%%%%%%%%%%%%%%%%%%%%%%
\maketitle
%\newpage
%
%%%%%%%%%%%%%%%%%%%%%%%%%%%%%%%%%%%%%%%%%
%\tableofcontents
%\section{Introduction}
%\label{sec:intro}

Higher temperature superconductors of correlated electron materials all display a `strange metal' phase above the critical temperature for superconductivity \cite{Hartnoll:2021ydi,syk_review}. This is a metallic phase of matter where the Landau quasi-particle approach breaks down. It is characterized most famously by a linear in temperature ($T$) electrical resistivity. We use the term strange metal only for those metals whose resistivity is smaller than the quantum unit ($h/e^2$ in $d=2$ spatial dimensions). Metals with a linear-in-$T$ resistivity which is larger than the quantum unit are `bad metals'. 

An often quoted model for a strange or bad metal ({\it e.g.\/} \cite{Sankar22,DasSarma24}) is one in which there is a large density of states of low energy bosonic excitations, usually phonons, and then quasi-elastic scattering of the electrons off the bosons leads to linear-in-$T$ resistivity from the Bose occupation function when $T$ is larger than the typical boson energy. However, studies of the optical conductivity in the strange metal of the cuprates \cite{Marel03} show that the dominant scattering is inelastic, not quasi-elastic, and leads to a non-Drude power-law-in-frequency tail in the optical conductivity. The optical conductivity data has been incisively analyzed recently by Michon {\it et al.} \cite{Michon23}: they have shown that while the transport scattering rate (related to the real part of the inverse optical conductivity) exhibits Planckian scaling behavior \cite{Hartnoll:2021ydi}, there are significant logarithmic deviations from scaling in the frequency and temperature dependent effective transport mass (related to the imaginary part of the inverse optical conductivity). Furthermore, the optical conductivity data connects consistently with d.c. measurements of resistivity and thermodynamics.

\begin{figure*}
    \centering
    \includegraphics{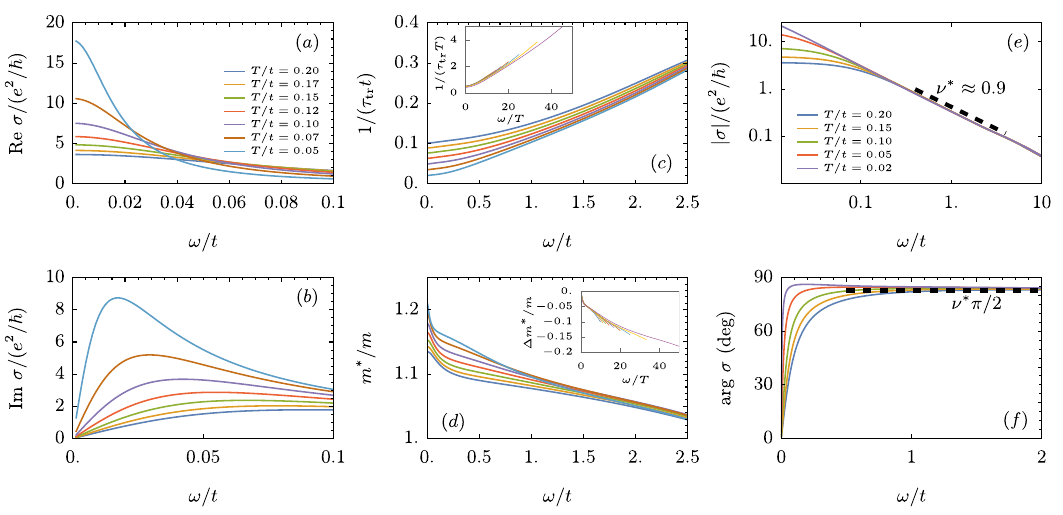}
    \caption{(a,b) Normal state optical conductivity at the quantum critical coupling for various temperatures indicated in Panel (a). (c,d) Transport scattering rate $1/\tau_{\rm tr}$ and effective mass $m^*$ obtained from $\sigma$ using (\ref{eq:Drude_formula}); inset of Panel (c) is a scaling plot with $1/\tau_{\rm tr}$ and $\omega$ scaled by $T$, inset of Panel (d) is a scaling plotting showing $\Delta m^* / m  = m^*(\omega)/m - m^*(0)/m$ as a function of $\omega/T$. (e,f) Modulus and phase of $\sigma(\omega)$. The effective exponent $\nu^*$ is explained in the text. Temperatures ranges are the same for Panels (a)-(d) and for Panels (e) and (f). Our results match the trends of the observations in Figs 3a,3b,1c,2b,1d,2d of Ref.~\onlinecite{Michon23}.}
    \label{fig:sig}
\end{figure*}

Our paper presents a self-consistent, disorder-averaged analysis of a two-dimensional Yukawa-Sachdev-Ye-Kitaev (2d-YSYK) model, which has a spatially random Yukawa coupling between fermions, $\psi$, with a Fermi surface and a 
nearly-critical scalar field, $\phi$.
We use methods similar to those which yield the exact solution of the zero-dimensional Sachdev-Ye-Kitaev model. Such a 2d-YSYK model has been argued \cite{Altman1,Patel2,PatelLunts,AAPQMC} to provide a universal description of quantum phase transitions in metals, associated with the condensation of $\phi$, in the presence of impurity-induced `Harris' disorder \cite{Harris,CCFS,Hoyos07} with spatial fluctuations in the local position of the quantum critical point. 
We find results that display all the key characteristics of the optical conductivity and d.c. resistivity described by Michon {\it et al.}, as shown in Fig.~\ref{fig:sig}.

\begin{figure*}
    \centering
    \includegraphics{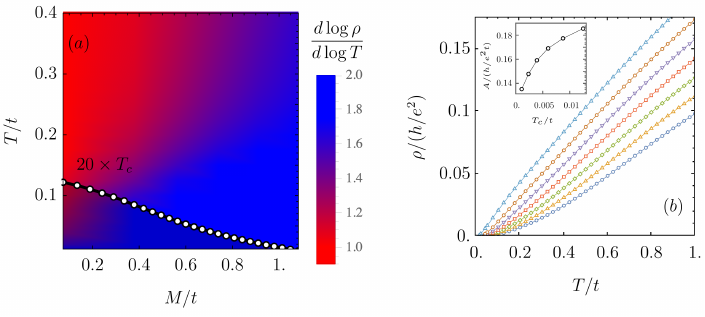}
    \caption{(a) Normal state resistivity exponent as a function of $T$ and $M$, the $T=0$ value of the renormalized boson mass, which tunes away from the quantum critical point at $M=0$ \cite{supp}, together with the superconducting $T_c$. Here the relatively small $T_c$ values have been multiplied to be put on the same scale as resistivity data. 
    (b) Normal state resistivity for different values of $M$. From bottom to top: $M/t = 1.3, 1.1, 0.9, 0.7, 0.6, 0.4, 0.$ The inset plots $A$, the co-efficient of the linear-$T$ resistivity, versus the superconducting $T_c$.}
    \label{fig:phasediag}
\end{figure*}

\begin{figure}
    \centering
    \includegraphics[width=\columnwidth]{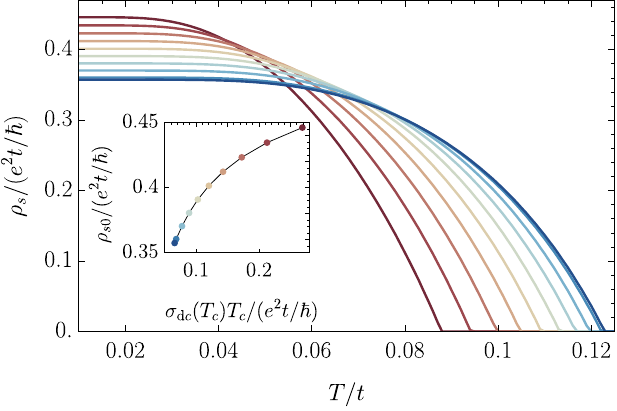}~~
    \caption{Temperature dependence of superfluid stiffness $\rho_s$ at $g'/t^{3/2}=5$ and $v=0$ for various values of renormalized boson mass $M$. Inset: The plot of zero temperature superfluid density $\rho_{s0}$ vs $\sigma_{\rm dc}(T_c)T_c$. From dark red to dark blue (critical point): $M/t = 1.7, 1.5, 1.4, 1.2, 1, 0.8, 0.7, 0.4, 0.1, 0.$}
    \label{fig:stiffness}
\end{figure}

Moreover, YSYK models also display instabilities of the strange metal to superconductivity \cite{Wang:2019bpd,Ilya1,WangMeng21,Schmalian2,Schmalian2bis}, with the pairing type dependent upon the particular quantum phase transition being studied. We examine an instability to spin-singlet pairing in a simplified model which ignores the gap variation around the Fermi surface, and therefore only applies to the anti-nodal regions of the cuprates.
We present the evolution of the electron spectral function and superfluid density for 
 $T < T_c$, the superconducting critical temperature.
We find that the 2d-YSYK model exhibits a number of experimentally observed trends:\\
({\it i\/}) It obeys the connection between scattering and pairing discussed by Taillefer \cite{Taillefer}, with a monotonic relation between $T_c$ and the slope of the linear-$T$ resistivity (Fig.~\ref{fig:phasediag}b).\\
({\it ii\/}) It has an overdoped regime where decreasing $T_c$ is accompanied by an increasing $T=0$ superfluid density (Fig.~\ref{fig:stiffness}a), as is observed in bulk samples of the cuprates \cite{Homes2}. Our interpretation is that the opening of a gap for the fermions $\psi$ in the superconducting state weakens the pairing interaction mediated by $\phi$ in a fully self-consistent theory at strong coupling \cite{Schmalian2,Schmalian2bis}. \\
({\it iii\/}) We study the relationship between the $T=0$ superfluid density and the normal state conductivity at $T_c$, and find a connection similar to Homes' Law \cite{Homes1} (Figs.~\ref{fig:stiffness}b and \ref{fig:stiffness_s} in the supplement \cite{supp}).

{\bf The 2d-YSYK Model}---Several works \cite{Hartnoll:2007ih,Maslov11,HMPS14,Patel14,MaslovChubukov17,Guo2022,Shi:2022toc,Schmalian1} have argued that clean quantum critical metals cannot serve as a universal model for transport in a strange metal, and that  impurity-induced spatial disorder is essential. We therefore add Harris disorder to the Hertz-Millis theory of a quantum phase transition in a metal associated with an Ising-nematic order parameter $\phi$ \cite{VojtaRMP}. Such disorder is provided by quenched random terms which preserve the Ising symmetry \cite{HMPS14,Patel14}. Other order parameters, including those at non-zero wavevector, and Fermi-volume changing transitions without broken symmetries \cite{SSORE}, also map to essentially the same 2d-YSYK model \cite{Patel2}. We define the 2d-YSYK model here as the Harris-disordered Hertz-Millis Lagrangian for $\phi$ and fermions $\psi$ with dispersion $\varepsilon ({\bm k})$ \cite{Altman1,Patel1,Patel2,Guo2022,Schmalian1,Ge:2024exw}:
\begin{align}
& \mathcal{L}_{\rm 2d-YSYK} = \psi^\dag_{i\sigma} \left[ \partial_\tau + \varepsilon (i\nabla) - \mu\right] \psi_{i\sigma} +    \frac{v_{ij}}{\sqrt{N}}({\bm r})\psi^\dag_{i\sigma}\psi_{i\sigma}\,  \nonumber \\
&~~+\frac 12  \left[ (\partial_\tau \phi_i)^2 + c^2 (\nabla \phi_i)^2 +s \phi_i^2 \right] +u  \phi_i^4 \nonumber  \\
&~~+ \frac{1}{N}[g_{ij\ell} +    g'_{ij\ell} ({\bm r})] \, \phi_{i\ell} \psi_{i\sigma}^\dag \mathcal D_{\bm r} \psi_{j\sigma}\,. \label{e9}
\end{align}
The flavor indices $i,j,\ell$ are summed over $N$ values. Imaginary time is $\tau$, we set $\hbar=1$, $s$ is the tuning parameter across the transition, $u$ is a scalar self-interaction. The operator ${\mathcal{D}}_{\bm r} = \partial_x^2 - \partial_y^2$ is special to the Ising-nematic case, and will be set to unity for simplicity in our computations as it unimportant apart from `cold spots' on the Fermi surface. In order to obtain a spin-singlet superconductor we have spinful fermions, with $\sigma$ the spin index. When the couplings are random in flavor space, the model is exactly solved at all $T$ in the large-$N$ limit by (\ref{eq:Yukawa_saddle}). Our discussion is for the physical case $N=1$, for which the saddle-point equations in (\ref{eq:Yukawa_saddle}) are applicable over an intermediate $T$ regime. 

$\mathcal{L}_{\rm 2d-YSYK}$ contains the two sources of disorder: One is the potential $v({\bm r})$ acting on the fermions. It is random in both flavor and space (in (\ref{vcorr}) and (\ref{gprimecorr}) we omit flavor indices for clarity, see appendix for full forms):
\be
\overline{v({\bm r})} = 0, \quad  \overline{v({\bm r}) v({\bm r'})} = v^2 \delta({\bm r}-{\bm r'}). \label{vcorr}
\ee
Its influence is familiar from the theory of weakly disordered metals, leading to marginally relevant localization effects on the fermions \cite{LeeRama}. 
Much more relevant is the Harris disorder, which we have taken in the form of a spatially random Yukawa
coupling $g' ({\bm r})$ adding to the spatially uniform Yukawa coupling $g$, and obeying 
\be
\overline{g'({\bm r})} = 0, \quad  \overline{g'({\bm r}) g'({\bm r'})} = g'^2 \delta({\bm r}-{\bm r'}). \label{gprimecorr}
\ee
By rescaling $\phi$, it is possible to transform $g' ({\bm r})$ into a more familiar `random mass' form of the Harris disorder, $s \rightarrow s + \delta s ({\bm r})$ \cite{Patel2}.
Both forms of the disorder have been examined in earlier work, random $g' ({\bm r})$ \cite{Patel2,AAPQMC} and random $\delta s ({\bm r})$ \cite{PatelLunts}, and similar results were obtained. 
We choose to work in the random $g' ({\bm r})$ formulation because it enables direct extension of methods employed for the exact solution of the zero-dimensional YSYK model \cite{Fu16,Murugan:2017eto,Patel:2018zpy,Marcus:2018tsr,Wang:2019bpd,Ilya1,Wang:2020dtj,WangMeng21,Schmalian3}. We restrict our analysis here to the simpler case with $g=0$ for then the self-energies are functions of frequency alone; earlier work \cite{Patel2} has shown that perturbative corrections in $g$ cancel in the transport response, partially justifying the $g=0$ choice. 
Two distinct regimes of behavior of $\mathcal{L}_{\rm 2d-YSYK}$ have been identified \cite{PatelLunts,AAPQMC}:\\
({\it i\/}) There is a significant intermediate energy regime where the bosonic and fermionic eigenmodes are spatially extended. The physics is self-averaging, and numerical results are consistent with the large-$N$ averaged Green's function methods which yield the exact solution of the zero-dimensional YSYK model. This is the regime we address in the present paper by standard SYK methods. The spatial disorder is needed for the singular single-particle self-energies to feed into singular transport properties \cite{Patel2}, which does not happen for translationally invariant models \cite{Hartnoll:2007ih,Maslov11,HMPS14,Patel14,MaslovChubukov17,Guo2022,Shi:2022toc,Schmalian1}. \\
({\it ii\/}) At low $T$ there is a crossover to a regime where spatial disorder causes bosonic eigenmodes to localize \cite{PatelLunts,AAPQMC}, while the fermionic eigenmodes remain extended \cite{MM}. Here, we must treat the disorder in the bosonic sector more completely: by a strong disorder renormalization \cite{Hoyos07}, numerically exactly \cite{PatelLunts,AAPQMC}, or map to `two-level system' models \cite{2LS23,2LS24}. Recent analyses \cite{Foster22,Grilli23} of $v({\bm r})$ disorder effects along the lines of Ref.~\onlinecite{LeeRama} near quantum criticality found singular corrections to the boson propagator, which we view as a precursor to boson localization. This boson localization regime is not described in the present paper.

%{\bf Equations for Green's functions}---
% can be obtained (see the supplement \cite{supp}) by endowing the fields with flavor indices, and making the Yukawa coupling also random in flavor space. After a disorder average, and in the limit of a large number of flavors where the saddle-point approximation is exact, we obtain the SYK-type equations which are indicated schematically in Fig.~\ref{fig:yukawa}.
%\begin{figure}[h]
%    \centering
%    \includegraphics[width=3in]{figure/yukawa.pdf}\\
%    \caption{Feynman diagrams for the $\psi$ self energy $\Sigma$, and the $\phi$ self energy $\Pi$. The wavy line is the $\phi$ Green's function, and the smooth line is the $\psi$ Green's function. All Green's functions include self energy corrections. The dashed line represents an average over spatial disorder. All Green's functions and self energies become $2 \times 2$ matrices in the superconducting phase. }
 %   \label{fig:yukawa}
%\end{figure}
{\bf Normal State}---We have solved the self-consistent equations for the Green's functions in (\ref{eq:Yukawa_saddle}) numerically on a 2D square lattice with nearest-neighbor hopping $t$ and with fermion chemical potential $\mu = -0.5 t$. The bosonic dispersion is chosen such that boson and fermion velocities are comparable $c\sim v_F$. These parameters are meant to represent the generic properties of our model. In the main text we primarily focus on results for two values of the interaction strength, $g'=2t^{3/2}$ and $g'=5 t^{3/2}$, which are representative weak coupling (interaction energy smaller than the fermion bandwidth) and strong coupling (interaction energy larger than the fermion bandwidth) values, respectively; we also focus on the case with no external impurity potential, $v=0$. See the supplement \cite{supp} for further details and results.
We summarize the approximate analytical solutions to (\ref{eq:Yukawa_saddle}) in the normal state, obtained earlier for a quadratic fermion band \cite{Patel2,Patel1}: the bosonic self-energy $\Pi(i\omega) \sim g'^2\mathcal N^2 |\omega|$, where $\mathcal N$ is the fermionic density of states, leading to overdamped, diffusive, bosonic dynamics at criticality with inverse boson Green's function $D^{-1}(q, i\omega) \sim |\omega| + \mathcal{D} q^2$ ($q$ is momentum); the fermion self-energy has a marginal Fermi-liquid \cite{Varma1989} form $\Sigma(i\omega) \sim i g'^2 \mathcal N \omega \log(|\omega|)$. The marginal Fermi liquid behavior does not extend to transport properties with a spatially uniform Yukawa coupling $g$ \cite{Patel2}, but does with the spatially random $g'$: the resistivity
was found to be $T$-linear up to logarithmic corrections, $\rho \sim g'^2 T \times \text{(logarithmic factors)}$.

Our numerical findings for the phase diagram are summarized in Fig.~\ref{fig:phasediag}a, which is plotted as a function of the renormalized boson mass $M$ used to tune the system to the quantum critical point (QCP), and $T$. 
The phenomenology is broadly similar to that observed experimentally in strange metals: Above the QCP there is a quantum-critical fan, in which the resistivity has an approximately linear $T$ dependence. At low $T$ the QCP is ultimately masked by a superconducting phase, with the maximal $T_c$ occurring at the critical point. 
The d.c. transport is shown in more detail in Fig.~\ref{fig:phasediag}b. Evidently, the logarithmic corrections to the resistivity mentioned above are relatively weak for the parameters we have considered. On the disordered side of the transition the resistivity is $T$-linear at elevated $T$ inside the quantum critical fan, before going to zero with with an approximately $T^2$ power-law below a certain crossover temperature. Over the entire $T$ range shown the resistivity is smaller than the quantum of resistance; the system is not a bad metal. 

We now consider the finite-frequency response \footnote{The related problem of optical conductivity at a \textit{clean} Ising-nematic QCP has been recently analyzed in \cite{li2023optical, gindikin2024fermi}.}. %The modulus and phase of the optical conductivity $\sigma(\omega)$ at the QCP are presented in Figs.~\ref{fig:phasediag}c,d. 
The real and imaginary parts of the optical conductivity are presented in Figs.~\ref{fig:sig}a and b. For the detailed structure encoded in $\sigma(\omega)$, we follow Michon \textit{et. al.} \cite{Michon23} and parametrize $\sigma(\omega)$ via a `generalized' Drude formula:
    \be
    \sigma(\omega) = i\frac{e^2 K/2}{\omega m^*(\omega)/m + i /\tau_\text{tr}(\omega)}. \label{eq:Drude_formula}
    \ee
Here $K$ is the optical weight and is equal to the average electronic kinetic energy (see (\ref{Kdef}) in the supplement \cite{supp}), allowing us to determine the frequency-dependent transport scattering rate $1/\tau_\text{tr}(\omega)$ and frequency dependent mass-enhancement parameter $m^*(\omega)/m$ directly from the data. 

The optical scattering rate is shown in Fig.~\ref{fig:sig}c. There is an approximately linear in frequency dependence down to $\omega \sim T$, along with $\omega/T$ scaling (see inset). For frequencies $\omega \lesssim T$, $1/\tau_\text{tr}$ tends to a $T$-dependent non-zero value which vanishes in the limit $T \to 0$. At larger $\omega$, the dimensionless ratio $1/[\omega \tau_{\rm tr} (\omega)]$ in Fig.~\ref{fig:sig}c is smaller than observed in \cite{Michon23}, but larger values of the ratio appear at larger $g'$ (see Fig.~\ref{fig:sig_g_5_inset}b \cite{supp}). 

The frequency-dependent mass enhancement parameter is shown in Fig.~\ref{fig:sig}d. For the chosen parameters, the low-frequency mass enhancement  does not exceed roughly 20\% down to the lowest temperatures we are able to reliably access in the numerical calculations. The behavior of this ``optical'' mass enhancement is consistent with the mass enhancement we have inferred from the fermion self-energy \cite{supp}. At sufficiently low $T$ $m^*(\omega = 0)/m$ is expected to diverge logarithmically with $T$ \cite{Patel2}. While we do not observe a pure logarithmic growth of $m^*/m$ in our data,  the mass enhancement does continuously increase with decreasing $T$ at the QCP, suggesting a (slow) divergence as $T \to 0$.

The modulus and phase of the optical conductivity are shown in Figs.~\ref{fig:sig}e and f: notably, we find the modulus has an apparent sub-linear power law behavior over an intermediate frequency range $|\sigma| \sim 1/\omega^{\nu^*}$, where $\nu^* \approx 0.9$ for the chosen set of parameters. Over a similar frequency range the phase of the optical conductivity plateaus at a value $\arg \sigma \approx \nu^* \pi /2$. 
Such behavior at intermediate frequencies has been observed in infrared conductivity measurements of cuprates \cite{Michon23,Marel03,norman2006high}. The exponent $\nu^*$ is continuously tunable with parameters and, in particular, is a decreasing function of the coupling strength $g'$ \cite{supp}. %This behavior at intermediate frequencies has been explained in Ref.~\cite{Michon23} on the basis of a general ``Planckian" model of the fermion self-energy, in which the scattering-rate is linear in energy. 

In the quantum-critical fan, $\omega/T$ scaling for the optical conductivity is spoiled by logarithmic corrections \cite{Patel2, Michon23}, the most significant such effect arising from the logarithmic divergence of the effective mass with $T$. This may be accounted for by utilizing the generalized Drude formula \eqref{eq:Drude_formula} and considering separately $\omega/T$ scaling of $1/\tau_\text{tr}$ and $\Delta m^*/m = m^*(\omega)/m - m^*(0)/m$, where the subtraction removes a contribution expected to violate scaling. With this approach we find reasonable scaling collapse for the optical scattering rate (inset of Fig.~\ref{fig:sig}c). For the parameters presented here, significant logarithmic corrections apparently remain for $\Delta m^*/m$. To the extent there is any reasonable scaling collapse, it only holds over a much narrow range of $\omega/T$ (inset of Fig.~\ref{fig:sig}d).

{\bf Superconductivity}---
The superconducting transition temperature $T_c$, as shown in Fig.~\ref{fig:phasediag}a, is numerically identified by the linearized gap equation; see the supplement \cite{supp}. 
For fixed $g'$, as we tune away from the QCP by increasing the renormalized boson mass $M$, we find that $T_c$ decreases. We compare $T_c$ with $A$, the slope of the linear-$T$ resistivity, in the inset of Fig.~\ref{fig:phasediag}b. The relation is monotonic, but not linear as discussed by Taillefer \cite{Taillefer}. We note, however, that in \cite{Taillefer} the coefficient $A$ was not extracted in the quantum-critical fan, which is how we have defined it, but rather from fitting to the low-$T$ resistivity in an ``extended" critical regime -- this regime has been associated with the localization of $\phi$ \cite{PatelLunts}, and is not addressed by our analysis here. If we remain at criticality while changing $g'$, then we do find a linear relationship between $A$ and $T_c$, as shown in Fig.~\ref{fig:Tc}a.

\begin{figure}
    \centering
    \includegraphics[width=\columnwidth]{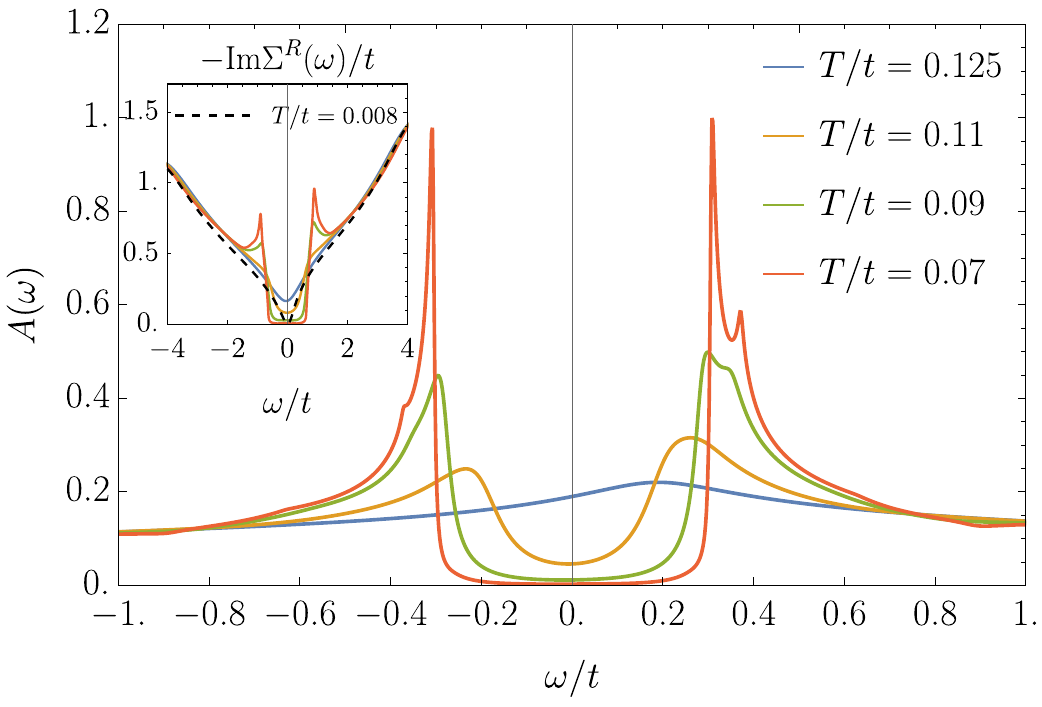}\\
    \caption{Evolution of the local density of states upon cooling through $T_c$ at the QCP for strong coupling $g'/t^{3/2}=5$. Inset shows the marginal Fermi-liquid form of the fermion self-energy $-\mathrm{Im} \Sigma \sim |\omega| $ at the QCP. Dashed curve is calculated in the normal state at $T/t=0.008$.}
    \label{fig:dos+sig}
\end{figure}

In Fig.~\ref{fig:dos+sig} we show the evolution of the electronic density of states at the QCP as the system goes through the superconducting transition, demonstrating an apparently conventional gap opening for $T<T_c$ despite the fact that, in the absence of superconductivity, the fermions behave as a marginal Fermi liquid with $-\mathrm{Im} \Sigma \sim |\omega|$ at low $T$, as seen in the inset of Fig. \ref{fig:dos+sig}. 

We have also computed the superfluid stiffness $\rho_s$ using (\ref{app:rhoK}). In Fig.~\ref{fig:stiffness}, we show the temperature dependence of $\rho_s$ for  $g'=5t^{3/2}$ as the system is tuned away from the QCP by varying the boson mass $M$. The stiffness reaches a finite constant value, $\rho_{s0}$, for $T\rightarrow 0$. As the system is tuned away from the QCP, $T_c$ {\it decreases\/} from its maximal value while $\rho_{s0}$ {\it increases\/}, as seen in Fig.~\ref{fig:stiffness}. This result at strong coupling aligns with the superfluid density in bulk samples of the overdoped cuprate superconductors \cite{Timusk09,Homes2}. When $T_c$ is largest at the QCP, the opening of a $\psi$ gap at $T=0$ weakens the pairing interaction mediated by $\phi$
(because the $\phi$ self energy $\Pi$ is self-consistently determined by the $\psi$ polarization, see Fig.~\ref{fig:yukawa}), leading to a smaller superfluid stiffness at $T=0$ \cite{Schmalian2,Schmalian2bis}.
On the other hand, in the presence of strong potential scattering $v$ this effect is weaker, and a decreasing $T_c$ is eventually accompanied by a decreasing $\rho_{s0}$, as shown in Fig.~\ref{fig:stiffness_s} \cite{supp}. 
The low $T$ boson localization \cite{PatelLunts} could have a significant influence on the spatial inhomogeneity of the superfluid density, and this remains to be studied.

Homes' Law \cite{Homes1} postulates a universal value for the dimensionless ratio $\rho_{s0}/(T_c \sigma_{\rm dc} (T_c))$, where $\sigma_{\rm dc}(T_c) $ is the normal state d.c. conductivity at $T_c$: we investigate this relationship in the inset of Fig.~\ref{fig:stiffness} and Section~\ref{app:homes} \cite{supp}. In the presence of a non-zero $v$, we find in Section~\ref{app:homes} a linear relation between $\rho_{s0}$ and $T_c \sigma_{\rm dc} (T_c)$, with a slope which can be 
close to the experimentally observed value.

{\bf Discussion}---The spatially inhomogeneous fermion-boson coupling $g' ({\bm r})$ in the 2d-YSYK model in \eqref{e9} is an alternative to `random mass' $\delta s ({\bm r})$ disorder in Hertz-Millis theory. The striking similarities between the results presented here and a variety of properties measured in strange metal superconductors indicate such 
%random-mass 
disorder plays a significant role in the phenomenology of these systems, making it important to determine the microscopic origin of disorder configurations that strongly affect the local position of the QCP. 
The enhancement of weak disorder by strong correlations \cite{PatelGeorges},
and the observation of spatial inhomogeneities in correlation-induced ordering \cite{Bianconi15}, are steps in this direction. 

{\bf Acknowledgements}---We thank Erez Berg, Christophe Berthod, Andrey Chubukov, Carlo Di Castro, Antoine Georges, Peter Lunts, Dmitrii Maslov, Srinivas Raghu, and Dirk van der Marel for valuable discussions.
This research was supported by the U.S. National Science Foundation grant No. DMR-2245246 and by the Simons Collaboration on Ultra-Quantum Matter which is a grant from the Simons Foundation (651440, S.S.). The Flatiron Institute is a division of the Simons Foundation. J.S. and D. V. were supported by the German Research Foundation (DFG) through CRC TRR 288 “ElastoQMat,” project A07. I.E. was supported by the University of Wisconsin–Madison.
\bibliography{Yukawa}

\begin{thebibliography}{58}%
\makeatletter
\providecommand \@ifxundefined [1]{%
 \@ifx{#1\undefined}
}%
\providecommand \@ifnum [1]{%
 \ifnum #1\expandafter \@firstoftwo
 \else \expandafter \@secondoftwo
 \fi
}%
\providecommand \@ifx [1]{%
 \ifx #1\expandafter \@firstoftwo
 \else \expandafter \@secondoftwo
 \fi
}%
\providecommand \natexlab [1]{#1}%
\providecommand \enquote  [1]{``#1''}%
\providecommand \bibnamefont  [1]{#1}%
\providecommand \bibfnamefont [1]{#1}%
\providecommand \citenamefont [1]{#1}%
\providecommand \href@noop [0]{\@secondoftwo}%
\providecommand \href [0]{\begingroup \@sanitize@url \@href}%
\providecommand \@href[1]{\@@startlink{#1}\@@href}%
\providecommand \@@href[1]{\endgroup#1\@@endlink}%
\providecommand \@sanitize@url [0]{\catcode `\\12\catcode `\$12\catcode `\&12\catcode `\#12\catcode `\^12\catcode `\_12\catcode `\%12\relax}%
\providecommand \@@startlink[1]{}%
\providecommand \@@endlink[0]{}%
\providecommand \url  [0]{\begingroup\@sanitize@url \@url }%
\providecommand \@url [1]{\endgroup\@href {#1}{\urlprefix }}%
\providecommand \urlprefix  [0]{URL }%
\providecommand \Eprint [0]{\href }%
\providecommand \doibase [0]{https://doi.org/}%
\providecommand \selectlanguage [0]{\@gobble}%
\providecommand \bibinfo  [0]{\@secondoftwo}%
\providecommand \bibfield  [0]{\@secondoftwo}%
\providecommand \translation [1]{[#1]}%
\providecommand \BibitemOpen [0]{}%
\providecommand \bibitemStop [0]{}%
\providecommand \bibitemNoStop [0]{.\EOS\space}%
\providecommand \EOS [0]{\spacefactor3000\relax}%
\providecommand \BibitemShut  [1]{\csname bibitem#1\endcsname}%
\let\auto@bib@innerbib\@empty
%</preamble>
\bibitem [{\citenamefont {Hartnoll}\ and\ \citenamefont {Mackenzie}(2022)}]{Hartnoll:2021ydi}%
  \BibitemOpen
  \bibfield  {author} {\bibinfo {author} {\bibfnamefont {S.~A.}\ \bibnamefont {Hartnoll}}\ and\ \bibinfo {author} {\bibfnamefont {A.~P.}\ \bibnamefont {Mackenzie}},\ }\bibfield  {title} {\emph {\bibinfo {title} {{Colloquium: Planckian dissipation in metals}}},\ }\href {https://doi.org/10.1103/RevModPhys.94.041002} {\bibfield  {journal} {\bibinfo  {journal} {Rev. Mod. Phys.}\ }\textbf {\bibinfo {volume} {94}},\ \bibinfo {pages} {041002} (\bibinfo {year} {2022})},\ \Eprint {https://arxiv.org/abs/2107.07802} {arXiv:2107.07802 [cond-mat.str-el]} \BibitemShut {NoStop}%
\bibitem [{\citenamefont {{Chowdhury}}\ \emph {et~al.}(2021)\citenamefont {{Chowdhury}}, \citenamefont {{Georges}}, \citenamefont {{Parcollet}},\ and\ \citenamefont {{Sachdev}}}]{syk_review}%
  \BibitemOpen
  \bibfield  {author} {\bibinfo {author} {\bibfnamefont {D.}~\bibnamefont {{Chowdhury}}}, \bibinfo {author} {\bibfnamefont {A.}~\bibnamefont {{Georges}}}, \bibinfo {author} {\bibfnamefont {O.}~\bibnamefont {{Parcollet}}},\ and\ \bibinfo {author} {\bibfnamefont {S.}~\bibnamefont {{Sachdev}}},\ }\bibfield  {title} {\emph {\bibinfo {title} {{Sachdev-Ye-Kitaev Models and Beyond: A Window into Non-Fermi Liquids}}},\ }\href@noop {} {\bibfield  {journal} {\bibinfo  {journal} {arXiv e-prints}\ ,\ \bibinfo {eid} {arXiv:2109.05037}} (\bibinfo {year} {2021})},\ \Eprint {https://arxiv.org/abs/2109.05037} {arXiv:2109.05037 [cond-mat.str-el]} \BibitemShut {NoStop}%
\bibitem [{\citenamefont {{Ahn}}\ and\ \citenamefont {{Das Sarma}}(2022)}]{Sankar22}%
  \BibitemOpen
  \bibfield  {author} {\bibinfo {author} {\bibfnamefont {S.}~\bibnamefont {{Ahn}}}\ and\ \bibinfo {author} {\bibfnamefont {S.}~\bibnamefont {{Das Sarma}}},\ }\bibfield  {title} {\emph {\bibinfo {title} {{Planckian properties of two-dimensional semiconductor systems}}},\ }\href {https://doi.org/10.1103/PhysRevB.106.155427} {\bibfield  {journal} {\bibinfo  {journal} {\prb}\ }\textbf {\bibinfo {volume} {106}},\ \bibinfo {eid} {155427} (\bibinfo {year} {2022})},\ \Eprint {https://arxiv.org/abs/2204.02982} {arXiv:2204.02982 [cond-mat.mes-hall]} \BibitemShut {NoStop}%
\bibitem [{\citenamefont {{Das Sarma}}\ and\ \citenamefont {{Tu}}(2024)}]{DasSarma24}%
  \BibitemOpen
  \bibfield  {author} {\bibinfo {author} {\bibfnamefont {S.}~\bibnamefont {{Das Sarma}}}\ and\ \bibinfo {author} {\bibfnamefont {Y.-T.}\ \bibnamefont {{Tu}}},\ }\bibfield  {title} {\emph {\bibinfo {title} {{Role of many-phonon modes on the high-temperature linear-in-T electronic resistivity}}},\ }\href {https://doi.org/10.1103/PhysRevB.109.235118} {\bibfield  {journal} {\bibinfo  {journal} {Phys. Rev. B}\ }\textbf {\bibinfo {volume} {109}},\ \bibinfo {eid} {235118} (\bibinfo {year} {2024})},\ \Eprint {https://arxiv.org/abs/2403.09890} {arXiv:2403.09890 [cond-mat.mes-hall]} \BibitemShut {NoStop}%
\bibitem [{\citenamefont {{Marel}}\ \emph {et~al.}(2003)\citenamefont {{Marel}}, \citenamefont {{Molegraaf}}, \citenamefont {{Zaanen}}, \citenamefont {{Nussinov}}, \citenamefont {{Carbone}}, \citenamefont {{Damascelli}}, \citenamefont {{Eisaki}}, \citenamefont {{Greven}}, \citenamefont {{Kes}},\ and\ \citenamefont {{Li}}}]{Marel03}%
  \BibitemOpen
  \bibfield  {author} {\bibinfo {author} {\bibfnamefont {D.~v.~d.}\ \bibnamefont {{Marel}}}, \bibinfo {author} {\bibfnamefont {H.~J.~A.}\ \bibnamefont {{Molegraaf}}}, \bibinfo {author} {\bibfnamefont {J.}~\bibnamefont {{Zaanen}}}, \bibinfo {author} {\bibfnamefont {Z.}~\bibnamefont {{Nussinov}}}, \bibinfo {author} {\bibfnamefont {F.}~\bibnamefont {{Carbone}}}, \bibinfo {author} {\bibfnamefont {A.}~\bibnamefont {{Damascelli}}}, \bibinfo {author} {\bibfnamefont {H.}~\bibnamefont {{Eisaki}}}, \bibinfo {author} {\bibfnamefont {M.}~\bibnamefont {{Greven}}}, \bibinfo {author} {\bibfnamefont {P.~H.}\ \bibnamefont {{Kes}}},\ and\ \bibinfo {author} {\bibfnamefont {M.}~\bibnamefont {{Li}}},\ }\bibfield  {title} {\emph {\bibinfo {title} {{Quantum critical behaviour in a high-T$_{c}$ superconductor}}},\ }\href {https://doi.org/10.1038/nature01978} {\bibfield  {journal} {\bibinfo  {journal} {\nat}\ }\textbf {\bibinfo {volume} {425}},\ \bibinfo {pages} {271} (\bibinfo {year} {2003})},\ \Eprint
  {https://arxiv.org/abs/cond-mat/0309172} {arXiv:cond-mat/0309172 [cond-mat.str-el]} \BibitemShut {NoStop}%
\bibitem [{\citenamefont {{Michon}}\ \emph {et~al.}(2023)\citenamefont {{Michon}}, \citenamefont {{Berthod}}, \citenamefont {{Rischau}}, \citenamefont {{Ataei}}, \citenamefont {{Chen}}, \citenamefont {{Komiya}}, \citenamefont {{Ono}}, \citenamefont {{Taillefer}}, \citenamefont {{van der Marel}},\ and\ \citenamefont {{Georges}}}]{Michon23}%
  \BibitemOpen
  \bibfield  {author} {\bibinfo {author} {\bibfnamefont {B.}~\bibnamefont {{Michon}}}, \bibinfo {author} {\bibfnamefont {C.}~\bibnamefont {{Berthod}}}, \bibinfo {author} {\bibfnamefont {C.~W.}\ \bibnamefont {{Rischau}}}, \bibinfo {author} {\bibfnamefont {A.}~\bibnamefont {{Ataei}}}, \bibinfo {author} {\bibfnamefont {L.}~\bibnamefont {{Chen}}}, \bibinfo {author} {\bibfnamefont {S.}~\bibnamefont {{Komiya}}}, \bibinfo {author} {\bibfnamefont {S.}~\bibnamefont {{Ono}}}, \bibinfo {author} {\bibfnamefont {L.}~\bibnamefont {{Taillefer}}}, \bibinfo {author} {\bibfnamefont {D.}~\bibnamefont {{van der Marel}}},\ and\ \bibinfo {author} {\bibfnamefont {A.}~\bibnamefont {{Georges}}},\ }\bibfield  {title} {\emph {\bibinfo {title} {{Reconciling scaling of the optical conductivity of cuprate superconductors with Planckian resistivity and specific heat}}},\ }\href {https://doi.org/10.1038/s41467-023-38762-5} {\bibfield  {journal} {\bibinfo  {journal} {Nature Communications}\ }\textbf {\bibinfo {volume} {14}},\ \bibinfo {eid}
  {3033} (\bibinfo {year} {2023})},\ \Eprint {https://arxiv.org/abs/2205.04030} {arXiv:2205.04030 [cond-mat.str-el]} \BibitemShut {NoStop}%
\bibitem [{\citenamefont {{Aldape}}\ \emph {et~al.}(2022)\citenamefont {{Aldape}}, \citenamefont {{Cookmeyer}}, \citenamefont {{Patel}},\ and\ \citenamefont {{Altman}}}]{Altman1}%
  \BibitemOpen
  \bibfield  {author} {\bibinfo {author} {\bibfnamefont {E.~E.}\ \bibnamefont {{Aldape}}}, \bibinfo {author} {\bibfnamefont {T.}~\bibnamefont {{Cookmeyer}}}, \bibinfo {author} {\bibfnamefont {A.~A.}\ \bibnamefont {{Patel}}},\ and\ \bibinfo {author} {\bibfnamefont {E.}~\bibnamefont {{Altman}}},\ }\bibfield  {title} {\emph {\bibinfo {title} {{Solvable theory of a strange metal at the breakdown of a heavy Fermi liquid}}},\ }\href {https://doi.org/10.1103/PhysRevB.105.235111} {\bibfield  {journal} {\bibinfo  {journal} {Phys. Rev. B}\ }\textbf {\bibinfo {volume} {105}},\ \bibinfo {eid} {235111} (\bibinfo {year} {2022})},\ \Eprint {https://arxiv.org/abs/2012.00763} {arXiv:2012.00763 [cond-mat.str-el]} \BibitemShut {NoStop}%
\bibitem [{\citenamefont {Patel}\ \emph {et~al.}(2023)\citenamefont {Patel}, \citenamefont {Guo}, \citenamefont {Esterlis},\ and\ \citenamefont {Sachdev}}]{Patel2}%
  \BibitemOpen
  \bibfield  {author} {\bibinfo {author} {\bibfnamefont {A.~A.}\ \bibnamefont {Patel}}, \bibinfo {author} {\bibfnamefont {H.}~\bibnamefont {Guo}}, \bibinfo {author} {\bibfnamefont {I.}~\bibnamefont {Esterlis}},\ and\ \bibinfo {author} {\bibfnamefont {S.}~\bibnamefont {Sachdev}},\ }\bibfield  {title} {\emph {\bibinfo {title} {{Universal theory of strange metals from spatially random interactions}}},\ }\href {https://doi.org/10.1126/science.abq6011} {\bibfield  {journal} {\bibinfo  {journal} {Science}\ }\textbf {\bibinfo {volume} {381}},\ \bibinfo {pages} {abq6011} (\bibinfo {year} {2023})},\ \Eprint {https://arxiv.org/abs/2203.04990} {arXiv:2203.04990 [cond-mat.str-el]} \BibitemShut {NoStop}%
\bibitem [{\citenamefont {{Patel}}\ \emph {et~al.}(2024{\natexlab{a}})\citenamefont {{Patel}}, \citenamefont {{Lunts}},\ and\ \citenamefont {{Sachdev}}}]{PatelLunts}%
  \BibitemOpen
  \bibfield  {author} {\bibinfo {author} {\bibfnamefont {A.~A.}\ \bibnamefont {{Patel}}}, \bibinfo {author} {\bibfnamefont {P.}~\bibnamefont {{Lunts}}},\ and\ \bibinfo {author} {\bibfnamefont {S.}~\bibnamefont {{Sachdev}}},\ }\bibfield  {title} {\emph {\bibinfo {title} {{Localization of overdamped bosonic modes and transport in strange metals}}},\ }\href {https://doi.org/10.1073/pnas.2402052121} {\bibfield  {journal} {\bibinfo  {journal} {Proc. Nat. Acad. Sci.}\ }\textbf {\bibinfo {volume} {121}},\ \bibinfo {pages} {e2402052121} (\bibinfo {year} {2024}{\natexlab{a}})},\ \Eprint {https://arxiv.org/abs/2312.06751} {arXiv:2312.06751 [cond-mat.str-el]} \BibitemShut {NoStop}%
\bibitem [{\citenamefont {{Patel}}\ \emph {et~al.}(2024{\natexlab{b}})\citenamefont {{Patel}}, \citenamefont {{Lunts}},\ and\ \citenamefont {{Albergo}}}]{AAPQMC}%
  \BibitemOpen
  \bibfield  {author} {\bibinfo {author} {\bibfnamefont {A.~A.}\ \bibnamefont {{Patel}}}, \bibinfo {author} {\bibfnamefont {P.}~\bibnamefont {{Lunts}}},\ and\ \bibinfo {author} {\bibfnamefont {M.}~\bibnamefont {{Albergo}}},\ }\bibfield  {title} {\emph {\bibinfo {title} {{Strange metals with spatially random interactions: a hybrid Monte Carlo study}}},\ }\href {https://www.icts.res.in/seminar/2024-07-26/aavishkar-patel} {\bibfield  {journal} {\bibinfo  {journal} {Talk at ICTS, Bengaluru}\ } (\bibinfo {year} {2024}{\natexlab{b}})}\BibitemShut {NoStop}%
\bibitem [{\citenamefont {Harris}(1974)}]{Harris}%
  \BibitemOpen
  \bibfield  {author} {\bibinfo {author} {\bibfnamefont {A.~B.}\ \bibnamefont {Harris}},\ }\bibfield  {title} {\emph {\bibinfo {title} {{Effect of random defects on the critical behaviour of Ising models}}},\ }\href {https://doi.org/10.1088/0022-3719/7/9/009} {\bibfield  {journal} {\bibinfo  {journal} {Journal of Physics C: Solid State Physics}\ }\textbf {\bibinfo {volume} {7}},\ \bibinfo {pages} {1671} (\bibinfo {year} {1974})}\BibitemShut {NoStop}%
\bibitem [{\citenamefont {Chayes}\ \emph {et~al.}(1986)\citenamefont {Chayes}, \citenamefont {Chayes}, \citenamefont {Fisher},\ and\ \citenamefont {Spencer}}]{CCFS}%
  \BibitemOpen
  \bibfield  {author} {\bibinfo {author} {\bibfnamefont {J.~T.}\ \bibnamefont {Chayes}}, \bibinfo {author} {\bibfnamefont {L.}~\bibnamefont {Chayes}}, \bibinfo {author} {\bibfnamefont {D.~S.}\ \bibnamefont {Fisher}},\ and\ \bibinfo {author} {\bibfnamefont {T.}~\bibnamefont {Spencer}},\ }\bibfield  {title} {\emph {\bibinfo {title} {{Finite-Size Scaling and Correlation Lengths for Disordered Systems}}},\ }\href {https://doi.org/10.1103/PhysRevLett.57.2999} {\bibfield  {journal} {\bibinfo  {journal} {Phys. Rev. Lett.}\ }\textbf {\bibinfo {volume} {57}},\ \bibinfo {pages} {2999} (\bibinfo {year} {1986})}\BibitemShut {NoStop}%
\bibitem [{\citenamefont {{Hoyos}}\ \emph {et~al.}(2007)\citenamefont {{Hoyos}}, \citenamefont {{Kotabage}},\ and\ \citenamefont {{Vojta}}}]{Hoyos07}%
  \BibitemOpen
  \bibfield  {author} {\bibinfo {author} {\bibfnamefont {J.~A.}\ \bibnamefont {{Hoyos}}}, \bibinfo {author} {\bibfnamefont {C.}~\bibnamefont {{Kotabage}}},\ and\ \bibinfo {author} {\bibfnamefont {T.}~\bibnamefont {{Vojta}}},\ }\bibfield  {title} {\emph {\bibinfo {title} {{Effects of Dissipation on a Quantum Critical Point with Disorder}}},\ }\href {https://doi.org/10.1103/PhysRevLett.99.230601} {\bibfield  {journal} {\bibinfo  {journal} {Phys. Rev. Lett.}\ }\textbf {\bibinfo {volume} {99}},\ \bibinfo {eid} {230601} (\bibinfo {year} {2007})},\ \Eprint {https://arxiv.org/abs/0705.1865} {arXiv:0705.1865 [cond-mat.str-el]} \BibitemShut {NoStop}%
\bibitem [{\citenamefont {Wang}(2020)}]{Wang:2019bpd}%
  \BibitemOpen
  \bibfield  {author} {\bibinfo {author} {\bibfnamefont {Y.}~\bibnamefont {Wang}},\ }\bibfield  {title} {\emph {\bibinfo {title} {{Solvable Strong-coupling Quantum Dot Model with a Non-Fermi-liquid Pairing Transition}}},\ }\href {https://doi.org/10.1103/PhysRevLett.124.017002} {\bibfield  {journal} {\bibinfo  {journal} {Phys. Rev. Lett.}\ }\textbf {\bibinfo {volume} {124}},\ \bibinfo {pages} {017002} (\bibinfo {year} {2020})},\ \Eprint {https://arxiv.org/abs/1904.07240} {arXiv:1904.07240 [cond-mat.str-el]} \BibitemShut {NoStop}%
\bibitem [{\citenamefont {Esterlis}\ and\ \citenamefont {Schmalian}(2019)}]{Ilya1}%
  \BibitemOpen
  \bibfield  {author} {\bibinfo {author} {\bibfnamefont {I.}~\bibnamefont {Esterlis}}\ and\ \bibinfo {author} {\bibfnamefont {J.}~\bibnamefont {Schmalian}},\ }\bibfield  {title} {\emph {\bibinfo {title} {{Cooper pairing of incoherent electrons: an electron-phonon version of the Sachdev-Ye-Kitaev model}}},\ }\href {https://doi.org/10.1103/PhysRevB.100.115132} {\bibfield  {journal} {\bibinfo  {journal} {Phys. Rev. B}\ }\textbf {\bibinfo {volume} {100}},\ \bibinfo {pages} {115132} (\bibinfo {year} {2019})},\ \Eprint {https://arxiv.org/abs/1906.04747} {arXiv:1906.04747 [cond-mat.str-el]} \BibitemShut {NoStop}%
\bibitem [{\citenamefont {{Wang}}\ \emph {et~al.}(2021)\citenamefont {{Wang}}, \citenamefont {{Davis}}, \citenamefont {{Pan}}, \citenamefont {{Wang}},\ and\ \citenamefont {{Meng}}}]{WangMeng21}%
  \BibitemOpen
  \bibfield  {author} {\bibinfo {author} {\bibfnamefont {W.}~\bibnamefont {{Wang}}}, \bibinfo {author} {\bibfnamefont {A.}~\bibnamefont {{Davis}}}, \bibinfo {author} {\bibfnamefont {G.}~\bibnamefont {{Pan}}}, \bibinfo {author} {\bibfnamefont {Y.}~\bibnamefont {{Wang}}},\ and\ \bibinfo {author} {\bibfnamefont {Z.~Y.}\ \bibnamefont {{Meng}}},\ }\bibfield  {title} {\emph {\bibinfo {title} {{Phase diagram of the spin-1/2 Yukawa-Sachdev-Ye-Kitaev model: Non-Fermi liquid, insulator, and superconductor}}},\ }\href {https://doi.org/10.1103/PhysRevB.103.195108} {\bibfield  {journal} {\bibinfo  {journal} {Phys. Rev. B}\ }\textbf {\bibinfo {volume} {103}},\ \bibinfo {eid} {195108} (\bibinfo {year} {2021})},\ \Eprint {https://arxiv.org/abs/2102.10755} {arXiv:2102.10755 [cond-mat.str-el]} \BibitemShut {NoStop}%
\bibitem [{\citenamefont {{Valentinis}}\ \emph {et~al.}(2023{\natexlab{a}})\citenamefont {{Valentinis}}, \citenamefont {{Inkof}},\ and\ \citenamefont {{Schmalian}}}]{Schmalian2}%
  \BibitemOpen
  \bibfield  {author} {\bibinfo {author} {\bibfnamefont {D.}~\bibnamefont {{Valentinis}}}, \bibinfo {author} {\bibfnamefont {G.~A.}\ \bibnamefont {{Inkof}}},\ and\ \bibinfo {author} {\bibfnamefont {J.}~\bibnamefont {{Schmalian}}},\ }\bibfield  {title} {\emph {\bibinfo {title} {{BCS to incoherent superconductivity crossovers in the Yukawa-SYK model on a lattice}}},\ }\href {https://doi.org/10.1103/PhysRevB.109.075162} {\bibfield  {journal} {\bibinfo  {journal} {Phys. Rev. B}\ }\textbf {\bibinfo {volume} {108}},\ \bibinfo {pages} {L140501} (\bibinfo {year} {2023}{\natexlab{a}})},\ \Eprint {https://arxiv.org/abs/2302.13138} {arXiv:2302.13138 [cond-mat.supr-con]} \BibitemShut {NoStop}%
\bibitem [{\citenamefont {{Valentinis}}\ \emph {et~al.}(2023{\natexlab{b}})\citenamefont {{Valentinis}}, \citenamefont {{Inkof}},\ and\ \citenamefont {{Schmalian}}}]{Schmalian2bis}%
  \BibitemOpen
  \bibfield  {author} {\bibinfo {author} {\bibfnamefont {D.}~\bibnamefont {{Valentinis}}}, \bibinfo {author} {\bibfnamefont {G.~A.}\ \bibnamefont {{Inkof}}},\ and\ \bibinfo {author} {\bibfnamefont {J.}~\bibnamefont {{Schmalian}}},\ }\bibfield  {title} {\emph {\bibinfo {title} {{Correlation between phase stiffness and condensation energy across the non-Fermi to Fermi-liquid crossover in the Yukawa-Sachdev-Ye-Kitaev model on a lattice}}},\ }\href {https://doi.org/10.1103/PhysRevResearch.5.043007} {\bibfield  {journal} {\bibinfo  {journal} {Phys. Rev. Res.}\ }\textbf {\bibinfo {volume} {5}},\ \bibinfo {pages} {043007} (\bibinfo {year} {2023}{\natexlab{b}})},\ \Eprint {https://arxiv.org/abs/2302.13134} {arXiv:2302.13134 [cond-mat.supr-con]} \BibitemShut {NoStop}%
\bibitem [{\citenamefont {{Taillefer}}(2010)}]{Taillefer}%
  \BibitemOpen
  \bibfield  {author} {\bibinfo {author} {\bibfnamefont {L.}~\bibnamefont {{Taillefer}}},\ }\bibfield  {title} {\emph {\bibinfo {title} {{Scattering and Pairing in Cuprate Superconductors}}},\ }\href {https://doi.org/10.1146/annurev-conmatphys-070909-104117} {\bibfield  {journal} {\bibinfo  {journal} {Annual Review of Condensed Matter Physics}\ }\textbf {\bibinfo {volume} {1}},\ \bibinfo {pages} {51} (\bibinfo {year} {2010})},\ \Eprint {https://arxiv.org/abs/1003.2972} {arXiv:1003.2972 [cond-mat.supr-con]} \BibitemShut {NoStop}%
\bibitem [{\citenamefont {{Dordevic}}\ and\ \citenamefont {{Homes}}(2022)}]{Homes2}%
  \BibitemOpen
  \bibfield  {author} {\bibinfo {author} {\bibfnamefont {S.~V.}\ \bibnamefont {{Dordevic}}}\ and\ \bibinfo {author} {\bibfnamefont {C.~C.}\ \bibnamefont {{Homes}}},\ }\bibfield  {title} {\emph {\bibinfo {title} {{Superfluid density in overdoped cuprates: Thin films versus bulk samples}}},\ }\href {https://doi.org/10.1103/PhysRevB.105.214514} {\bibfield  {journal} {\bibinfo  {journal} {\prb}\ }\textbf {\bibinfo {volume} {105}},\ \bibinfo {eid} {214514} (\bibinfo {year} {2022})},\ \Eprint {https://arxiv.org/abs/2206.13987} {arXiv:2206.13987 [cond-mat.supr-con]} \BibitemShut {NoStop}%
\bibitem [{\citenamefont {{Homes}}\ \emph {et~al.}(2004)\citenamefont {{Homes}}, \citenamefont {{Dordevic}}, \citenamefont {{Strongin}}, \citenamefont {{Bonn}}, \citenamefont {{Liang}}, \citenamefont {{Hardy}}, \citenamefont {{Komiya}}, \citenamefont {{Ando}}, \citenamefont {{Yu}}, \citenamefont {{Kaneko}}, \citenamefont {{Zhao}}, \citenamefont {{Greven}}, \citenamefont {{Basov}},\ and\ \citenamefont {{Timusk}}}]{Homes1}%
  \BibitemOpen
  \bibfield  {author} {\bibinfo {author} {\bibfnamefont {C.~C.}\ \bibnamefont {{Homes}}}, \bibinfo {author} {\bibfnamefont {S.~V.}\ \bibnamefont {{Dordevic}}}, \bibinfo {author} {\bibfnamefont {M.}~\bibnamefont {{Strongin}}}, \bibinfo {author} {\bibfnamefont {D.~A.}\ \bibnamefont {{Bonn}}}, \bibinfo {author} {\bibfnamefont {R.}~\bibnamefont {{Liang}}}, \bibinfo {author} {\bibfnamefont {W.~N.}\ \bibnamefont {{Hardy}}}, \bibinfo {author} {\bibfnamefont {S.}~\bibnamefont {{Komiya}}}, \bibinfo {author} {\bibfnamefont {Y.}~\bibnamefont {{Ando}}}, \bibinfo {author} {\bibfnamefont {G.}~\bibnamefont {{Yu}}}, \bibinfo {author} {\bibfnamefont {N.}~\bibnamefont {{Kaneko}}}, \bibinfo {author} {\bibfnamefont {X.}~\bibnamefont {{Zhao}}}, \bibinfo {author} {\bibfnamefont {M.}~\bibnamefont {{Greven}}}, \bibinfo {author} {\bibfnamefont {D.~N.}\ \bibnamefont {{Basov}}},\ and\ \bibinfo {author} {\bibfnamefont {T.}~\bibnamefont {{Timusk}}},\ }\bibfield  {title} {\emph {\bibinfo {title} {{A universal scaling relation in
  high-temperature superconductors}}},\ }\href {https://doi.org/10.1038/nature02673} {\bibfield  {journal} {\bibinfo  {journal} {\nat}\ }\textbf {\bibinfo {volume} {430}},\ \bibinfo {pages} {539} (\bibinfo {year} {2004})},\ \Eprint {https://arxiv.org/abs/cond-mat/0404216} {arXiv:cond-mat/0404216 [cond-mat.supr-con]} \BibitemShut {NoStop}%
\bibitem [{sup()}]{supp}%
  \BibitemOpen
  \bibinfo {note} {See {Supplemental Material}, which includes {Refs.~\cite{Ilya1,Hauck,FFT,schmalian1996self,Patel1,Michon23,Scalapino93,Homes1}}, for additional data and more detailed derivations}\BibitemShut {NoStop}%
\bibitem [{\citenamefont {Hartnoll}\ \emph {et~al.}(2007)\citenamefont {Hartnoll}, \citenamefont {Kovtun}, \citenamefont {Muller},\ and\ \citenamefont {Sachdev}}]{Hartnoll:2007ih}%
  \BibitemOpen
  \bibfield  {author} {\bibinfo {author} {\bibfnamefont {S.~A.}\ \bibnamefont {Hartnoll}}, \bibinfo {author} {\bibfnamefont {P.~K.}\ \bibnamefont {Kovtun}}, \bibinfo {author} {\bibfnamefont {M.}~\bibnamefont {Muller}},\ and\ \bibinfo {author} {\bibfnamefont {S.}~\bibnamefont {Sachdev}},\ }\bibfield  {title} {\emph {\bibinfo {title} {{Theory of the Nernst effect near quantum phase transitions in condensed matter, and in dyonic black holes}}},\ }\href {https://doi.org/10.1103/PhysRevB.76.144502} {\bibfield  {journal} {\bibinfo  {journal} {Phys. Rev. B}\ }\textbf {\bibinfo {volume} {76}},\ \bibinfo {pages} {144502} (\bibinfo {year} {2007})},\ \Eprint {https://arxiv.org/abs/0706.3215} {arXiv:0706.3215 [cond-mat.str-el]} \BibitemShut {NoStop}%
\bibitem [{\citenamefont {{Maslov}}\ \emph {et~al.}(2011)\citenamefont {{Maslov}}, \citenamefont {{Yudson}},\ and\ \citenamefont {{Chubukov}}}]{Maslov11}%
  \BibitemOpen
  \bibfield  {author} {\bibinfo {author} {\bibfnamefont {D.~L.}\ \bibnamefont {{Maslov}}}, \bibinfo {author} {\bibfnamefont {V.~I.}\ \bibnamefont {{Yudson}}},\ and\ \bibinfo {author} {\bibfnamefont {A.~V.}\ \bibnamefont {{Chubukov}}},\ }\bibfield  {title} {\emph {\bibinfo {title} {{Resistivity of a Non-Galilean-Invariant Fermi Liquid near Pomeranchuk Quantum Criticality}}},\ }\href {https://doi.org/10.1103/PhysRevLett.106.106403} {\bibfield  {journal} {\bibinfo  {journal} {\prl}\ }\textbf {\bibinfo {volume} {106}},\ \bibinfo {eid} {106403} (\bibinfo {year} {2011})},\ \Eprint {https://arxiv.org/abs/1012.0069} {arXiv:1012.0069 [cond-mat.str-el]} \BibitemShut {NoStop}%
\bibitem [{\citenamefont {{Hartnoll}}\ \emph {et~al.}(2014)\citenamefont {{Hartnoll}}, \citenamefont {{Mahajan}}, \citenamefont {{Punk}},\ and\ \citenamefont {{Sachdev}}}]{HMPS14}%
  \BibitemOpen
  \bibfield  {author} {\bibinfo {author} {\bibfnamefont {S.~A.}\ \bibnamefont {{Hartnoll}}}, \bibinfo {author} {\bibfnamefont {R.}~\bibnamefont {{Mahajan}}}, \bibinfo {author} {\bibfnamefont {M.}~\bibnamefont {{Punk}}},\ and\ \bibinfo {author} {\bibfnamefont {S.}~\bibnamefont {{Sachdev}}},\ }\bibfield  {title} {\emph {\bibinfo {title} {{Transport near the Ising-nematic quantum critical point of metals in two dimensions}}},\ }\href {https://doi.org/10.1103/PhysRevB.89.155130} {\bibfield  {journal} {\bibinfo  {journal} {Phys. Rev. B}\ }\textbf {\bibinfo {volume} {89}},\ \bibinfo {eid} {155130} (\bibinfo {year} {2014})},\ \Eprint {https://arxiv.org/abs/1401.7012} {arXiv:1401.7012 [cond-mat.str-el]} \BibitemShut {NoStop}%
\bibitem [{\citenamefont {{Patel}}\ and\ \citenamefont {{Sachdev}}(2014)}]{Patel14}%
  \BibitemOpen
  \bibfield  {author} {\bibinfo {author} {\bibfnamefont {A.~A.}\ \bibnamefont {{Patel}}}\ and\ \bibinfo {author} {\bibfnamefont {S.}~\bibnamefont {{Sachdev}}},\ }\bibfield  {title} {\emph {\bibinfo {title} {{dc resistivity at the onset of spin density wave order in two-dimensional metals}}},\ }\href {https://doi.org/10.1103/PhysRevB.90.165146} {\bibfield  {journal} {\bibinfo  {journal} {Phys. Rev. B}\ }\textbf {\bibinfo {volume} {90}},\ \bibinfo {eid} {165146} (\bibinfo {year} {2014})},\ \Eprint {https://arxiv.org/abs/1408.6549} {arXiv:1408.6549 [cond-mat.str-el]} \BibitemShut {NoStop}%
\bibitem [{\citenamefont {{Maslov}}\ and\ \citenamefont {{Chubukov}}(2017)}]{MaslovChubukov17}%
  \BibitemOpen
  \bibfield  {author} {\bibinfo {author} {\bibfnamefont {D.~L.}\ \bibnamefont {{Maslov}}}\ and\ \bibinfo {author} {\bibfnamefont {A.~V.}\ \bibnamefont {{Chubukov}}},\ }\bibfield  {title} {\emph {\bibinfo {title} {{Optical response of correlated electron systems}}},\ }\href {https://doi.org/10.1088/1361-6633/80/2/026503} {\bibfield  {journal} {\bibinfo  {journal} {Reports on Progress in Physics}\ }\textbf {\bibinfo {volume} {80}},\ \bibinfo {eid} {026503} (\bibinfo {year} {2017})},\ \Eprint {https://arxiv.org/abs/1608.02514} {arXiv:1608.02514 [cond-mat.str-el]} \BibitemShut {NoStop}%
\bibitem [{\citenamefont {Guo}\ \emph {et~al.}(2022)\citenamefont {Guo}, \citenamefont {Patel}, \citenamefont {Esterlis},\ and\ \citenamefont {Sachdev}}]{Guo2022}%
  \BibitemOpen
  \bibfield  {author} {\bibinfo {author} {\bibfnamefont {H.}~\bibnamefont {Guo}}, \bibinfo {author} {\bibfnamefont {A.~A.}\ \bibnamefont {Patel}}, \bibinfo {author} {\bibfnamefont {I.}~\bibnamefont {Esterlis}},\ and\ \bibinfo {author} {\bibfnamefont {S.}~\bibnamefont {Sachdev}},\ }\bibfield  {title} {\emph {\bibinfo {title} {{Large-$N$ theory of critical Fermi surfaces. II. Conductivity}}},\ }\href {https://doi.org/10.1103/PhysRevB.106.115151} {\bibfield  {journal} {\bibinfo  {journal} {Phys. Rev. B}\ }\textbf {\bibinfo {volume} {106}},\ \bibinfo {pages} {115151} (\bibinfo {year} {2022})},\ \Eprint {https://arxiv.org/abs/2207.08841} {arXiv:2207.08841 [cond-mat.str-el]} \BibitemShut {NoStop}%
\bibitem [{\citenamefont {Shi}\ \emph {et~al.}(2023)\citenamefont {Shi}, \citenamefont {Else}, \citenamefont {Goldman},\ and\ \citenamefont {Senthil}}]{Shi:2022toc}%
  \BibitemOpen
  \bibfield  {author} {\bibinfo {author} {\bibfnamefont {Z.~D.}\ \bibnamefont {Shi}}, \bibinfo {author} {\bibfnamefont {D.~V.}\ \bibnamefont {Else}}, \bibinfo {author} {\bibfnamefont {H.}~\bibnamefont {Goldman}},\ and\ \bibinfo {author} {\bibfnamefont {T.}~\bibnamefont {Senthil}},\ }\bibfield  {title} {\emph {\bibinfo {title} {{Loop current fluctuations and quantum critical transport}}},\ }\href {https://doi.org/10.21468/SciPostPhys.14.5.113} {\bibfield  {journal} {\bibinfo  {journal} {SciPost Phys.}\ }\textbf {\bibinfo {volume} {14}},\ \bibinfo {pages} {113} (\bibinfo {year} {2023})},\ \Eprint {https://arxiv.org/abs/2208.04328} {arXiv:2208.04328 [cond-mat.str-el]} \BibitemShut {NoStop}%
\bibitem [{\citenamefont {Guo}\ \emph {et~al.}(2024)\citenamefont {Guo}, \citenamefont {Valentinis}, \citenamefont {Schmalian}, \citenamefont {Sachdev},\ and\ \citenamefont {Patel}}]{Schmalian1}%
  \BibitemOpen
  \bibfield  {author} {\bibinfo {author} {\bibfnamefont {H.}~\bibnamefont {Guo}}, \bibinfo {author} {\bibfnamefont {D.}~\bibnamefont {Valentinis}}, \bibinfo {author} {\bibfnamefont {J.}~\bibnamefont {Schmalian}}, \bibinfo {author} {\bibfnamefont {S.}~\bibnamefont {Sachdev}},\ and\ \bibinfo {author} {\bibfnamefont {A.~A.}\ \bibnamefont {Patel}},\ }\bibfield  {title} {\emph {\bibinfo {title} {{Cyclotron resonance and quantum oscillations of critical Fermi surfaces}}},\ }\href {https://doi.org/10.1103/PhysRevB.109.075162} {\bibfield  {journal} {\bibinfo  {journal} {Phys. Rev. B}\ }\textbf {\bibinfo {volume} {109}},\ \bibinfo {pages} {075162} (\bibinfo {year} {2024})},\ \Eprint {https://arxiv.org/abs/:2308.01956} {arXiv::2308.01956 [cond-mat.supr-con]} \BibitemShut {NoStop}%
\bibitem [{\citenamefont {{L{\"o}hneysen}}\ \emph {et~al.}(2007)\citenamefont {{L{\"o}hneysen}}, \citenamefont {{Rosch}}, \citenamefont {{Vojta}},\ and\ \citenamefont {{W{\"o}lfle}}}]{VojtaRMP}%
  \BibitemOpen
  \bibfield  {author} {\bibinfo {author} {\bibfnamefont {H.~V.}\ \bibnamefont {{L{\"o}hneysen}}}, \bibinfo {author} {\bibfnamefont {A.}~\bibnamefont {{Rosch}}}, \bibinfo {author} {\bibfnamefont {M.}~\bibnamefont {{Vojta}}},\ and\ \bibinfo {author} {\bibfnamefont {P.}~\bibnamefont {{W{\"o}lfle}}},\ }\bibfield  {title} {\emph {\bibinfo {title} {{Fermi-liquid instabilities at magnetic quantum phase transitions}}},\ }\href {https://doi.org/10.1103/RevModPhys.79.1015} {\bibfield  {journal} {\bibinfo  {journal} {Rev. Mod. Phys.}\ }\textbf {\bibinfo {volume} {79}},\ \bibinfo {pages} {1015} (\bibinfo {year} {2007})},\ \Eprint {https://arxiv.org/abs/cond-mat/0606317} {arXiv:cond-mat/0606317 [cond-mat.str-el]} \BibitemShut {NoStop}%
\bibitem [{\citenamefont {Sachdev}(2023)}]{SSORE}%
  \BibitemOpen
  \bibfield  {author} {\bibinfo {author} {\bibfnamefont {S.}~\bibnamefont {Sachdev}},\ }\bibfield  {title} {\emph {\bibinfo {title} {{Strange metals and black holes: insights from the Sachdev-Ye-Kitaev model}}},\ }\bibfield  {journal} {\bibinfo  {journal} {Oxford Research Encyclopedia in Physics}\ }\href {https://doi.org/10.1093/acrefore/9780190871994.013.48} {10.1093/acrefore/9780190871994.013.48} (\bibinfo {year} {2023}),\ \Eprint {https://arxiv.org/abs/2305.01001} {arXiv:2305.01001 [cond-mat.str-el]} \BibitemShut {NoStop}%
\bibitem [{\citenamefont {Esterlis}\ \emph {et~al.}(2021)\citenamefont {Esterlis}, \citenamefont {Guo}, \citenamefont {Patel},\ and\ \citenamefont {Sachdev}}]{Patel1}%
  \BibitemOpen
  \bibfield  {author} {\bibinfo {author} {\bibfnamefont {I.}~\bibnamefont {Esterlis}}, \bibinfo {author} {\bibfnamefont {H.}~\bibnamefont {Guo}}, \bibinfo {author} {\bibfnamefont {A.~A.}\ \bibnamefont {Patel}},\ and\ \bibinfo {author} {\bibfnamefont {S.}~\bibnamefont {Sachdev}},\ }\bibfield  {title} {\emph {\bibinfo {title} {{Large $N$ theory of critical Fermi surfaces}}},\ }\href {https://doi.org/10.1103/PhysRevB.103.235129} {\bibfield  {journal} {\bibinfo  {journal} {Phys. Rev. B}\ }\textbf {\bibinfo {volume} {103}},\ \bibinfo {pages} {235129} (\bibinfo {year} {2021})},\ \Eprint {https://arxiv.org/abs/2103.08615} {arXiv:2103.08615 [cond-mat.str-el]} \BibitemShut {NoStop}%
\bibitem [{\citenamefont {Ge}\ \emph {et~al.}(2024)\citenamefont {Ge}, \citenamefont {Sin},\ and\ \citenamefont {Wang}}]{Ge:2024exw}%
  \BibitemOpen
  \bibfield  {author} {\bibinfo {author} {\bibfnamefont {X.-H.}\ \bibnamefont {Ge}}, \bibinfo {author} {\bibfnamefont {S.-J.}\ \bibnamefont {Sin}},\ and\ \bibinfo {author} {\bibfnamefont {Y.-L.}\ \bibnamefont {Wang}},\ }\bibfield  {title} {\emph {\bibinfo {title} {{Linear-T Resistivity from Spatially Random Vector Coupling}}},\ }\href@noop {} {\  (\bibinfo {year} {2024})},\ \Eprint {https://arxiv.org/abs/2406.11170} {arXiv:2406.11170 [hep-th]} \BibitemShut {NoStop}%
\bibitem [{\citenamefont {Lee}\ and\ \citenamefont {Ramakrishnan}(1985)}]{LeeRama}%
  \BibitemOpen
  \bibfield  {author} {\bibinfo {author} {\bibfnamefont {P.~A.}\ \bibnamefont {Lee}}\ and\ \bibinfo {author} {\bibfnamefont {T.~V.}\ \bibnamefont {Ramakrishnan}},\ }\bibfield  {title} {\emph {\bibinfo {title} {Disordered electronic systems}},\ }\href {https://doi.org/10.1103/RevModPhys.57.287} {\bibfield  {journal} {\bibinfo  {journal} {Rev. Mod. Phys.}\ }\textbf {\bibinfo {volume} {57}},\ \bibinfo {pages} {287} (\bibinfo {year} {1985})}\BibitemShut {NoStop}%
\bibitem [{\citenamefont {Fu}\ \emph {et~al.}(2017)\citenamefont {Fu}, \citenamefont {Gaiotto}, \citenamefont {Maldacena},\ and\ \citenamefont {Sachdev}}]{Fu16}%
  \BibitemOpen
  \bibfield  {author} {\bibinfo {author} {\bibfnamefont {W.}~\bibnamefont {Fu}}, \bibinfo {author} {\bibfnamefont {D.}~\bibnamefont {Gaiotto}}, \bibinfo {author} {\bibfnamefont {J.}~\bibnamefont {Maldacena}},\ and\ \bibinfo {author} {\bibfnamefont {S.}~\bibnamefont {Sachdev}},\ }\bibfield  {title} {\emph {\bibinfo {title} {{Supersymmetric Sachdev-Ye-Kitaev models}}},\ }\href {https://doi.org/10.1103/PhysRevD.95.026009} {\bibfield  {journal} {\bibinfo  {journal} {Phys. Rev. D}\ }\textbf {\bibinfo {volume} {95}},\ \bibinfo {pages} {026009} (\bibinfo {year} {2017})},\ \bibinfo {note} {[Addendum: Phys.Rev.D 95, 069904 (2017)]},\ \Eprint {https://arxiv.org/abs/1610.08917} {arXiv:1610.08917 [hep-th]} \BibitemShut {NoStop}%
\bibitem [{\citenamefont {Murugan}\ \emph {et~al.}(2017)\citenamefont {Murugan}, \citenamefont {Stanford},\ and\ \citenamefont {Witten}}]{Murugan:2017eto}%
  \BibitemOpen
  \bibfield  {author} {\bibinfo {author} {\bibfnamefont {J.}~\bibnamefont {Murugan}}, \bibinfo {author} {\bibfnamefont {D.}~\bibnamefont {Stanford}},\ and\ \bibinfo {author} {\bibfnamefont {E.}~\bibnamefont {Witten}},\ }\bibfield  {title} {\emph {\bibinfo {title} {{More on Supersymmetric and 2d Analogs of the SYK Model}}},\ }\href {https://doi.org/10.1007/JHEP08(2017)146} {\bibfield  {journal} {\bibinfo  {journal} {JHEP}\ }\textbf {\bibinfo {volume} {08}},\ \bibinfo {pages} {146}},\ \Eprint {https://arxiv.org/abs/1706.05362} {arXiv:1706.05362 [hep-th]} \BibitemShut {NoStop}%
\bibitem [{\citenamefont {Patel}\ and\ \citenamefont {Sachdev}(2018)}]{Patel:2018zpy}%
  \BibitemOpen
  \bibfield  {author} {\bibinfo {author} {\bibfnamefont {A.~A.}\ \bibnamefont {Patel}}\ and\ \bibinfo {author} {\bibfnamefont {S.}~\bibnamefont {Sachdev}},\ }\bibfield  {title} {\emph {\bibinfo {title} {{Critical strange metal from fluctuating gauge fields in a solvable random model}}},\ }\href {https://doi.org/10.1103/PhysRevB.98.125134} {\bibfield  {journal} {\bibinfo  {journal} {Phys. Rev. B}\ }\textbf {\bibinfo {volume} {98}},\ \bibinfo {pages} {125134} (\bibinfo {year} {2018})},\ \Eprint {https://arxiv.org/abs/1807.04754} {arXiv:1807.04754 [cond-mat.str-el]} \BibitemShut {NoStop}%
\bibitem [{\citenamefont {Marcus}\ and\ \citenamefont {Vandoren}(2019)}]{Marcus:2018tsr}%
  \BibitemOpen
  \bibfield  {author} {\bibinfo {author} {\bibfnamefont {E.}~\bibnamefont {Marcus}}\ and\ \bibinfo {author} {\bibfnamefont {S.}~\bibnamefont {Vandoren}},\ }\bibfield  {title} {\emph {\bibinfo {title} {{A new class of SYK-like models with maximal chaos}}},\ }\href {https://doi.org/10.1007/JHEP01(2019)166} {\bibfield  {journal} {\bibinfo  {journal} {JHEP}\ }\textbf {\bibinfo {volume} {01}},\ \bibinfo {pages} {166}},\ \Eprint {https://arxiv.org/abs/1808.01190} {arXiv:1808.01190 [hep-th]} \BibitemShut {NoStop}%
\bibitem [{\citenamefont {Wang}\ and\ \citenamefont {Chubukov}(2020)}]{Wang:2020dtj}%
  \BibitemOpen
  \bibfield  {author} {\bibinfo {author} {\bibfnamefont {Y.}~\bibnamefont {Wang}}\ and\ \bibinfo {author} {\bibfnamefont {A.~V.}\ \bibnamefont {Chubukov}},\ }\bibfield  {title} {\emph {\bibinfo {title} {{Quantum Phase Transition in the Yukawa-SYK Model}}},\ }\href {https://doi.org/10.1103/PhysRevResearch.2.033084} {\bibfield  {journal} {\bibinfo  {journal} {Phys. Rev. Res.}\ }\textbf {\bibinfo {volume} {2}},\ \bibinfo {pages} {033084} (\bibinfo {year} {2020})},\ \Eprint {https://arxiv.org/abs/2005.07205} {arXiv:2005.07205 [cond-mat.str-el]} \BibitemShut {NoStop}%
\bibitem [{\citenamefont {{Hosseinabadi}}\ \emph {et~al.}(2023)\citenamefont {{Hosseinabadi}}, \citenamefont {{Kelly}}, \citenamefont {{Schmalian}},\ and\ \citenamefont {{Marino}}}]{Schmalian3}%
  \BibitemOpen
  \bibfield  {author} {\bibinfo {author} {\bibfnamefont {H.}~\bibnamefont {{Hosseinabadi}}}, \bibinfo {author} {\bibfnamefont {S.~P.}\ \bibnamefont {{Kelly}}}, \bibinfo {author} {\bibfnamefont {J.}~\bibnamefont {{Schmalian}}},\ and\ \bibinfo {author} {\bibfnamefont {J.}~\bibnamefont {{Marino}}},\ }\bibfield  {title} {\emph {\bibinfo {title} {{Thermalization of non-Fermi-liquid electron-phonon systems: Hydrodynamic relaxation of the Yukawa-Sachdev-Ye-Kitaev model}}},\ }\href {https://doi.org/10.1103/PhysRevB.108.104319} {\bibfield  {journal} {\bibinfo  {journal} {Phys. Rev. B}\ }\textbf {\bibinfo {volume} {108}},\ \bibinfo {eid} {104319} (\bibinfo {year} {2023})},\ \Eprint {https://arxiv.org/abs/2306.03898} {arXiv:2306.03898 [cond-mat.str-el]} \BibitemShut {NoStop}%
\bibitem [{\citenamefont {Milovanovi\ifmmode~\acute{c}\else \'{c}\fi{}}\ \emph {et~al.}(1989)\citenamefont {Milovanovi\ifmmode~\acute{c}\else \'{c}\fi{}}, \citenamefont {Sachdev},\ and\ \citenamefont {Bhatt}}]{MM}%
  \BibitemOpen
  \bibfield  {author} {\bibinfo {author} {\bibfnamefont {M.}~\bibnamefont {Milovanovi\ifmmode~\acute{c}\else \'{c}\fi{}}}, \bibinfo {author} {\bibfnamefont {S.}~\bibnamefont {Sachdev}},\ and\ \bibinfo {author} {\bibfnamefont {R.~N.}\ \bibnamefont {Bhatt}},\ }\bibfield  {title} {\emph {\bibinfo {title} {Effective-field theory of local-moment formation in disordered metals}},\ }\href {https://doi.org/10.1103/PhysRevLett.63.82} {\bibfield  {journal} {\bibinfo  {journal} {Phys. Rev. Lett.}\ }\textbf {\bibinfo {volume} {63}},\ \bibinfo {pages} {82} (\bibinfo {year} {1989})}\BibitemShut {NoStop}%
\bibitem [{\citenamefont {Bashan}\ \emph {et~al.}(2024)\citenamefont {Bashan}, \citenamefont {Tulipman}, \citenamefont {Schmalian},\ and\ \citenamefont {Berg}}]{2LS23}%
  \BibitemOpen
  \bibfield  {author} {\bibinfo {author} {\bibfnamefont {N.}~\bibnamefont {Bashan}}, \bibinfo {author} {\bibfnamefont {E.}~\bibnamefont {Tulipman}}, \bibinfo {author} {\bibfnamefont {J.}~\bibnamefont {Schmalian}},\ and\ \bibinfo {author} {\bibfnamefont {E.}~\bibnamefont {Berg}},\ }\bibfield  {title} {\emph {\bibinfo {title} {{Tunable Non-Fermi Liquid Phase from Coupling to Two-Level Systems}}},\ }\href {https://doi.org/10.1103/PhysRevLett.132.236501} {\bibfield  {journal} {\bibinfo  {journal} {Phys. Rev. Lett.}\ }\textbf {\bibinfo {volume} {132}},\ \bibinfo {pages} {236501} (\bibinfo {year} {2024})},\ \Eprint {https://arxiv.org/abs/2310.07768} {arXiv:2310.07768 [cond-mat.str-el]} \BibitemShut {NoStop}%
\bibitem [{\citenamefont {{Tulipman}}\ \emph {et~al.}(2024)\citenamefont {{Tulipman}}, \citenamefont {{Bashan}}, \citenamefont {{Schmalian}},\ and\ \citenamefont {{Berg}}}]{2LS24}%
  \BibitemOpen
  \bibfield  {author} {\bibinfo {author} {\bibfnamefont {E.}~\bibnamefont {{Tulipman}}}, \bibinfo {author} {\bibfnamefont {N.}~\bibnamefont {{Bashan}}}, \bibinfo {author} {\bibfnamefont {J.}~\bibnamefont {{Schmalian}}},\ and\ \bibinfo {author} {\bibfnamefont {E.}~\bibnamefont {{Berg}}},\ }\bibfield  {title} {\emph {\bibinfo {title} {{Solvable models of two-level systems coupled to itinerant electrons: Robust non-Fermi liquid and quantum critical pairing}}},\ }\href@noop {} {\  (\bibinfo {year} {2024})},\ \Eprint {https://arxiv.org/abs/2404.06532} {arXiv:2404.06532 [cond-mat.str-el]} \BibitemShut {NoStop}%
\bibitem [{\citenamefont {{Wu}}\ \emph {et~al.}(2022)\citenamefont {{Wu}}, \citenamefont {{Liao}},\ and\ \citenamefont {{Foster}}}]{Foster22}%
  \BibitemOpen
  \bibfield  {author} {\bibinfo {author} {\bibfnamefont {T.~C.}\ \bibnamefont {{Wu}}}, \bibinfo {author} {\bibfnamefont {Y.}~\bibnamefont {{Liao}}},\ and\ \bibinfo {author} {\bibfnamefont {M.~S.}\ \bibnamefont {{Foster}}},\ }\bibfield  {title} {\emph {\bibinfo {title} {{Quantum interference of hydrodynamic modes in a dirty marginal Fermi liquid}}},\ }\href {https://doi.org/10.1103/PhysRevB.106.155108} {\bibfield  {journal} {\bibinfo  {journal} {Phys. Rev. B}\ }\textbf {\bibinfo {volume} {106}},\ \bibinfo {eid} {155108} (\bibinfo {year} {2022})},\ \Eprint {https://arxiv.org/abs/2206.01762} {arXiv:2206.01762 [cond-mat.str-el]} \BibitemShut {NoStop}%
\bibitem [{\citenamefont {{Grilli}}\ \emph {et~al.}(2023)\citenamefont {{Grilli}}, \citenamefont {{Di Castro}}, \citenamefont {{Mirarchi}}, \citenamefont {{Seibold}},\ and\ \citenamefont {{Caprara}}}]{Grilli23}%
  \BibitemOpen
  \bibfield  {author} {\bibinfo {author} {\bibfnamefont {M.}~\bibnamefont {{Grilli}}}, \bibinfo {author} {\bibfnamefont {C.}~\bibnamefont {{Di Castro}}}, \bibinfo {author} {\bibfnamefont {G.}~\bibnamefont {{Mirarchi}}}, \bibinfo {author} {\bibfnamefont {G.}~\bibnamefont {{Seibold}}},\ and\ \bibinfo {author} {\bibfnamefont {S.}~\bibnamefont {{Caprara}}},\ }\bibfield  {title} {\emph {\bibinfo {title} {Dissipative quantum criticality as a source of strange metal behavior}},\ }\href {https://doi.org/10.3390/sym15030569} {\bibfield  {journal} {\bibinfo  {journal} {Symmetry}\ }\textbf {\bibinfo {volume} {15}},\ \bibinfo {pages} {569} (\bibinfo {year} {2023})},\ \Eprint {https://arxiv.org/abs/2205.10876} {arXiv:2205.10876 [cond-mat.str-el]} \BibitemShut {NoStop}%
\bibitem [{\citenamefont {{Varma}}\ \emph {et~al.}(1989)\citenamefont {{Varma}}, \citenamefont {{Littlewood}}, \citenamefont {{Schmitt-Rink}}, \citenamefont {{Abrahams}},\ and\ \citenamefont {{Ruckenstein}}}]{Varma1989}%
  \BibitemOpen
  \bibfield  {author} {\bibinfo {author} {\bibfnamefont {C.~M.}\ \bibnamefont {{Varma}}}, \bibinfo {author} {\bibfnamefont {P.~B.}\ \bibnamefont {{Littlewood}}}, \bibinfo {author} {\bibfnamefont {S.}~\bibnamefont {{Schmitt-Rink}}}, \bibinfo {author} {\bibfnamefont {E.}~\bibnamefont {{Abrahams}}},\ and\ \bibinfo {author} {\bibfnamefont {A.~E.}\ \bibnamefont {{Ruckenstein}}},\ }\bibfield  {title} {\emph {\bibinfo {title} {{Phenomenology of the normal state of Cu-O high-temperature superconductors}}},\ }\href {https://doi.org/10.1103/PhysRevLett.63.1996} {\bibfield  {journal} {\bibinfo  {journal} {\prl}\ }\textbf {\bibinfo {volume} {63}},\ \bibinfo {pages} {1996} (\bibinfo {year} {1989})}\BibitemShut {NoStop}%
\bibitem [{Note1()}]{Note1}%
  \BibitemOpen
  \bibinfo {note} {The related problem of optical conductivity at a \protect \textit {clean} Ising-nematic QCP has been recently analyzed in \cite {li2023optical, gindikin2024fermi}.}\BibitemShut {Stop}%
\bibitem [{\citenamefont {Norman}\ and\ \citenamefont {Chubukov}(2006)}]{norman2006high}%
  \BibitemOpen
  \bibfield  {author} {\bibinfo {author} {\bibfnamefont {M.~R.}\ \bibnamefont {Norman}}\ and\ \bibinfo {author} {\bibfnamefont {A.~V.}\ \bibnamefont {Chubukov}},\ }\bibfield  {title} {\emph {\bibinfo {title} {High-frequency behavior of the infrared conductivity of cuprates}},\ }\href {https://doi.org/10.1103/PhysRevB.73.140501} {\bibfield  {journal} {\bibinfo  {journal} {Phys. Rev. B}\ }\textbf {\bibinfo {volume} {73}},\ \bibinfo {pages} {140501} (\bibinfo {year} {2006})}\BibitemShut {NoStop}%
\bibitem [{\citenamefont {{Hwang}}\ \emph {et~al.}(2007)\citenamefont {{Hwang}}, \citenamefont {{Timusk}},\ and\ \citenamefont {{Gu}}}]{Timusk09}%
  \BibitemOpen
  \bibfield  {author} {\bibinfo {author} {\bibfnamefont {J.}~\bibnamefont {{Hwang}}}, \bibinfo {author} {\bibfnamefont {T.}~\bibnamefont {{Timusk}}},\ and\ \bibinfo {author} {\bibfnamefont {G.~D.}\ \bibnamefont {{Gu}}},\ }\bibfield  {title} {\emph {\bibinfo {title} {{Doping dependent optical properties of Bi$_{2}$Sr$_{2}$CaCu$_{2}$O$_{8+{\ensuremath{\delta}}}$}}},\ }\href {https://doi.org/10.1088/0953-8984/19/12/125208} {\bibfield  {journal} {\bibinfo  {journal} {Journal of Physics Condensed Matter}\ }\textbf {\bibinfo {volume} {19}},\ \bibinfo {eid} {125208} (\bibinfo {year} {2007})},\ \Eprint {https://arxiv.org/abs/cond-mat/0607653} {arXiv:cond-mat/0607653 [cond-mat.supr-con]} \BibitemShut {NoStop}%
\bibitem [{\citenamefont {{Hardy}}\ \emph {et~al.}(2024)\citenamefont {{Hardy}}, \citenamefont {{Parcollet}}, \citenamefont {{Georges}},\ and\ \citenamefont {{Patel}}}]{PatelGeorges}%
  \BibitemOpen
  \bibfield  {author} {\bibinfo {author} {\bibfnamefont {A.}~\bibnamefont {{Hardy}}}, \bibinfo {author} {\bibfnamefont {O.}~\bibnamefont {{Parcollet}}}, \bibinfo {author} {\bibfnamefont {A.}~\bibnamefont {{Georges}}},\ and\ \bibinfo {author} {\bibfnamefont {A.~A.}\ \bibnamefont {{Patel}}},\ }\bibfield  {title} {\emph {\bibinfo {title} {{Enhanced strange metallicity from Coulomb repulsion}}},\ }\href@noop {} {\bibfield  {journal} {\bibinfo  {journal} {arXiv e-prints}\ } (\bibinfo {year} {2024})},\ \Eprint {https://arxiv.org/abs/2407.21102} {arXiv:2407.21102 [cond-mat.str-el]} \BibitemShut {NoStop}%
\bibitem [{\citenamefont {{Campi}}\ \emph {et~al.}(2015)\citenamefont {{Campi}}, \citenamefont {{Bianconi}}, \citenamefont {{Poccia}}, \citenamefont {{Bianconi}}, \citenamefont {{Barba}}, \citenamefont {{Arrighetti}}, \citenamefont {{Innocenti}}, \citenamefont {{Karpinski}}, \citenamefont {{Zhigadlo}}, \citenamefont {{Kazakov}}, \citenamefont {{Burghammer}}, \citenamefont {{Zimmermann}}, \citenamefont {{Sprung}},\ and\ \citenamefont {{Ricci}}}]{Bianconi15}%
  \BibitemOpen
  \bibfield  {author} {\bibinfo {author} {\bibfnamefont {G.}~\bibnamefont {{Campi}}}, \bibinfo {author} {\bibfnamefont {A.}~\bibnamefont {{Bianconi}}}, \bibinfo {author} {\bibfnamefont {N.}~\bibnamefont {{Poccia}}}, \bibinfo {author} {\bibfnamefont {G.}~\bibnamefont {{Bianconi}}}, \bibinfo {author} {\bibfnamefont {L.}~\bibnamefont {{Barba}}}, \bibinfo {author} {\bibfnamefont {G.}~\bibnamefont {{Arrighetti}}}, \bibinfo {author} {\bibfnamefont {D.}~\bibnamefont {{Innocenti}}}, \bibinfo {author} {\bibfnamefont {J.}~\bibnamefont {{Karpinski}}}, \bibinfo {author} {\bibfnamefont {N.~D.}\ \bibnamefont {{Zhigadlo}}}, \bibinfo {author} {\bibfnamefont {S.~M.}\ \bibnamefont {{Kazakov}}}, \bibinfo {author} {\bibfnamefont {M.}~\bibnamefont {{Burghammer}}}, \bibinfo {author} {\bibfnamefont {M.~V.}\ \bibnamefont {{Zimmermann}}}, \bibinfo {author} {\bibfnamefont {M.}~\bibnamefont {{Sprung}}},\ and\ \bibinfo {author} {\bibfnamefont {A.}~\bibnamefont {{Ricci}}},\ }\bibfield  {title} {\emph {\bibinfo {title} {{Inhomogeneity of
  charge-density-wave order and quenched disorder in a high-T$_{c}$ superconductor}}},\ }\href {https://doi.org/10.1038/nature14987} {\bibfield  {journal} {\bibinfo  {journal} {Nature}\ }\textbf {\bibinfo {volume} {525}},\ \bibinfo {pages} {359} (\bibinfo {year} {2015})},\ \Eprint {https://arxiv.org/abs/1509.05002} {arXiv:1509.05002 [cond-mat.supr-con]} \BibitemShut {NoStop}%
\bibitem [{\citenamefont {Hauck}\ \emph {et~al.}(2020)\citenamefont {Hauck}, \citenamefont {Klug}, \citenamefont {Esterlis},\ and\ \citenamefont {Schmalian}}]{Hauck}%
  \BibitemOpen
  \bibfield  {author} {\bibinfo {author} {\bibfnamefont {D.}~\bibnamefont {Hauck}}, \bibinfo {author} {\bibfnamefont {M.~J.}\ \bibnamefont {Klug}}, \bibinfo {author} {\bibfnamefont {I.}~\bibnamefont {Esterlis}},\ and\ \bibinfo {author} {\bibfnamefont {J.}~\bibnamefont {Schmalian}},\ }\bibfield  {title} {\emph {\bibinfo {title} {{Eliashberg equations for an electron–phonon version of the Sachdev–Ye–Kitaev model: Pair breaking in non-Fermi liquid superconductors}}},\ }\href {https://doi.org/https://doi.org/10.1016/j.aop.2020.168120} {\bibfield  {journal} {\bibinfo  {journal} {Annals of Physics}\ }\textbf {\bibinfo {volume} {417}},\ \bibinfo {pages} {168120} (\bibinfo {year} {2020})}\BibitemShut {NoStop}%
\bibitem [{\citenamefont {Dee}\ \emph {et~al.}(2019)\citenamefont {Dee}, \citenamefont {Nakatsukasa}, \citenamefont {Wang},\ and\ \citenamefont {Johnston}}]{FFT}%
  \BibitemOpen
  \bibfield  {author} {\bibinfo {author} {\bibfnamefont {P.~M.}\ \bibnamefont {Dee}}, \bibinfo {author} {\bibfnamefont {K.}~\bibnamefont {Nakatsukasa}}, \bibinfo {author} {\bibfnamefont {Y.}~\bibnamefont {Wang}},\ and\ \bibinfo {author} {\bibfnamefont {S.}~\bibnamefont {Johnston}},\ }\bibfield  {title} {\emph {\bibinfo {title} {{Temperature-filling phase diagram of the two-dimensional Holstein model in the thermodynamic limit by self-consistent Migdal approximation}}},\ }\href {https://doi.org/10.1103/PhysRevB.99.024514} {\bibfield  {journal} {\bibinfo  {journal} {Phys. Rev. B}\ }\textbf {\bibinfo {volume} {99}},\ \bibinfo {pages} {024514} (\bibinfo {year} {2019})}\BibitemShut {NoStop}%
\bibitem [{\citenamefont {Schmalian}\ \emph {et~al.}(1996)\citenamefont {Schmalian}, \citenamefont {Langer}, \citenamefont {Grabowski},\ and\ \citenamefont {Bennemann}}]{schmalian1996self}%
  \BibitemOpen
  \bibfield  {author} {\bibinfo {author} {\bibfnamefont {J.}~\bibnamefont {Schmalian}}, \bibinfo {author} {\bibfnamefont {M.}~\bibnamefont {Langer}}, \bibinfo {author} {\bibfnamefont {S.}~\bibnamefont {Grabowski}},\ and\ \bibinfo {author} {\bibfnamefont {K.}~\bibnamefont {Bennemann}},\ }\bibfield  {title} {\emph {\bibinfo {title} {Self-consistent summation of many-particle diagrams on the real frequency axis and its application to the flex approximation}},\ }\href@noop {} {\bibfield  {journal} {\bibinfo  {journal} {Computer physics communications}\ }\textbf {\bibinfo {volume} {93}},\ \bibinfo {pages} {141} (\bibinfo {year} {1996})}\BibitemShut {NoStop}%
\bibitem [{\citenamefont {Scalapino}\ \emph {et~al.}(1993)\citenamefont {Scalapino}, \citenamefont {White},\ and\ \citenamefont {Zhang}}]{Scalapino93}%
  \BibitemOpen
  \bibfield  {author} {\bibinfo {author} {\bibfnamefont {D.~J.}\ \bibnamefont {Scalapino}}, \bibinfo {author} {\bibfnamefont {S.~R.}\ \bibnamefont {White}},\ and\ \bibinfo {author} {\bibfnamefont {S.}~\bibnamefont {Zhang}},\ }\bibfield  {title} {\emph {\bibinfo {title} {Insulator, metal, or superconductor: The criteria}},\ }\href {https://doi.org/10.1103/PhysRevB.47.7995} {\bibfield  {journal} {\bibinfo  {journal} {Phys. Rev. B}\ }\textbf {\bibinfo {volume} {47}},\ \bibinfo {pages} {7995} (\bibinfo {year} {1993})}\BibitemShut {NoStop}%
\bibitem [{\citenamefont {{Li}}\ \emph {et~al.}(2023)\citenamefont {{Li}}, \citenamefont {{Sharma}}, \citenamefont {{Levchenko}},\ and\ \citenamefont {{Maslov}}}]{li2023optical}%
  \BibitemOpen
  \bibfield  {author} {\bibinfo {author} {\bibfnamefont {S.}~\bibnamefont {{Li}}}, \bibinfo {author} {\bibfnamefont {P.}~\bibnamefont {{Sharma}}}, \bibinfo {author} {\bibfnamefont {A.}~\bibnamefont {{Levchenko}}},\ and\ \bibinfo {author} {\bibfnamefont {D.~L.}\ \bibnamefont {{Maslov}}},\ }\bibfield  {title} {\emph {\bibinfo {title} {{Optical conductivity of a metal near an Ising-nematic quantum critical point}}},\ }\href {https://doi.org/10.1103/PhysRevB.108.235125} {\bibfield  {journal} {\bibinfo  {journal} {\prb}\ }\textbf {\bibinfo {volume} {108}},\ \bibinfo {eid} {235125} (\bibinfo {year} {2023})},\ \Eprint {https://arxiv.org/abs/2309.12571} {arXiv:2309.12571 [cond-mat.str-el]} \BibitemShut {NoStop}%
\bibitem [{\citenamefont {{Gindikin}}\ and\ \citenamefont {{Chubukov}}(2024)}]{gindikin2024fermi}%
  \BibitemOpen
  \bibfield  {author} {\bibinfo {author} {\bibfnamefont {Y.}~\bibnamefont {{Gindikin}}}\ and\ \bibinfo {author} {\bibfnamefont {A.~V.}\ \bibnamefont {{Chubukov}}},\ }\bibfield  {title} {\emph {\bibinfo {title} {{Fermi surface geometry and optical conductivity of a two-dimensional electron gas near an Ising-nematic quantum critical point}}},\ }\href {https://doi.org/10.1103/PhysRevB.109.115156} {\bibfield  {journal} {\bibinfo  {journal} {\prb}\ }\textbf {\bibinfo {volume} {109}},\ \bibinfo {eid} {115156} (\bibinfo {year} {2024})},\ \Eprint {https://arxiv.org/abs/2401.17392} {arXiv:2401.17392 [cond-mat.str-el]} \BibitemShut {NoStop}%
\end{thebibliography}%

\onecolumngrid
\begin{center}
{\bf\large End Matter}
\end{center}
%\twocolumngrid

\appendix*
 \setcounter{equation}{0}

{\it Appendix: The 2\lowercase{d}-YSYK model and its saddle-point equations} -- We work with the following lattice  action in imaginary time:
%\begin{widetext}
\begin{align}
S& = \int \rd\tau\rd^2 x  \sum_{i=1}^N\sum_{\sigma=\pm 1}\psi^\dagger_{i\sigma}(\tau,x)\left[\partial_\tau+\varepsilon_k\right]\psi_{i\sigma}(\tau,x)
+\frac{1}{2}\int \rd\tau\rd^2 x \sum_{i=1}^N \phi_{i}(\tau,x)\left[-\partial_\tau^2+\om_q^2\right]\phi_{i}(\tau,x) \nn
&+ \int \rd^2 \tau \rd x \sum_{i,j}^N\sum_{\sigma=\pm 1} \frac{v_{ij} (x) }{\sqrt{N}}\psi_{i\sigma}^{\dagger} (\tau,x) \psi_{j\sigma}( \tau,x) +\int \rd\tau\rd^2 x \sum_{i,j,\ell=1}^N\sum_{\sigma=\pm 1}\frac{g_{ij\ell}^{'}(x)}{N}\psi^\dagger_{i\sigma}(\tau,x)\psi_{j\sigma}(\tau,x)\phi_{\ell}(\tau,x)\,,
%+\int \rd\tau\rd^2 x \sum_{i,j=1}^N\sum_\sigma\left[\frac{v_{ij}(x)}{\sqrt{N}}\psi^\dagger_{i\sigma}(\tau,x)\psi_{j\sigma}(\tau,x)\right],
\label{eq:latticeaction}
\end{align}
%\end{widetext}
where $i,j,\ell$ are flavor indices and $\sigma$ is the spin index. We use the lattice dispersions
\begin{subequations}
   \begin{align}
    \epsilon_k  &= -2t (\cos k_x  + \cos k_y)-\mu \,,\\
    \omega^2_q  &= s + 2J(2- \cos q_x - \cos q_y)\,.
\end{align} 
 \label{ep:disp}%
\end{subequations}
Here $t$ is the fermion hopping, $\mu$ is the chemical potential, $s$ is the (squared) bare boson mass and the stiffness $J$ determines the boson dispersion. The Yukawa couplings $g_{ijl}^{'}$ are complex-valued random variables that obey $g_{ijl}^{'}(x)=g_{1,ij\ell} (x) + i g_{2,ij\ell} (x) = g_{ji\ell}^{'}(x)$. The real part $g_{1,ij\ell}(x)$ and the imaginary part $g_{2,ij\ell}(x)$ have zero mean and the variances are \cite{Hauck}
\begin{eqnarray}
\overline{g_{1,ij\ell}(x)g_{1,i'j'\ell'}(x')} & = & \left(1-\frac{\alpha}{2}\right)g'^2\delta_{\ell,\ell'}\left(\delta_{ii'}\delta_{jj'}+\delta_{ij'}\delta_{ji'}\right)\delta(x-x'),\nonumber \\
\overline{g_{2,ij\ell}(x)g_{2,i'j'\ell'}(x')} & = & \frac{\alpha}{2}g'^2\delta_{\ell,\ell'}\left(\delta_{ii'}\delta_{jj'}-\delta_{ij'}\delta_{ji'}\right)\delta(x-x'),\nonumber \\
\overline{g_{1,ij\ell}(x')g_{2,i'j',\ell'}(x')} & = & 0.\label{eq:distribution}
\end{eqnarray}
In the $\alpha=1$ limit, \eqref{eq:distribution} reduces to $\overline{g_{ij\ell}^{'}(x) g_{i'j'\ell'}^{'}(x')^*}=g'^2\delta(x-x')\delta_{ii'}\delta_{jj'}\delta_{kk'}$ and no superconductivity occurs \cite{Hauck}. For $\alpha=0$, however, the coupling constants are all real valued, which preserves time-reversal symmetry for each realization of the random couplings and thus gives rise to  superconductivity. However, note that the saddle-point equations in (\ref{eq:Yukawa_saddle}) below are independent of $\alpha$ in the normal state.
The spatially random potential satisfies
\begin{align}
    & \overline{v_{ij} (x)} = 0 \,, \quad \overline{v^\ast_{ij} (x) v_{i'j'} (x')} = v^2 \, \delta(x-x') \delta_{ii'} \delta_{jj'}\,,
    \label{eq:vdistribution}
\end{align} 
To tune the system to criticality, we have found it most convenient to first impose a fixed length constraint \cite{Patel1}:  
\begin{equation}\label{eq:constraint}
    \sum_q\sum_{i=1}^N \phi_{iq}(\tau)\phi_{i,-q}(\tau)=N/\gamma\,.
\end{equation}
This equation implicitly determines $s$, using $\gamma$ as tuning parameter to access the QCP. In the main text we present the phase diagram in terms of the $T=0$ value of the renormalized boson mass $M^2 = s - \Pi(\omega = 0, T=0)$.

\begin{figure}[h]
    \centering
    \includegraphics[width=5in]{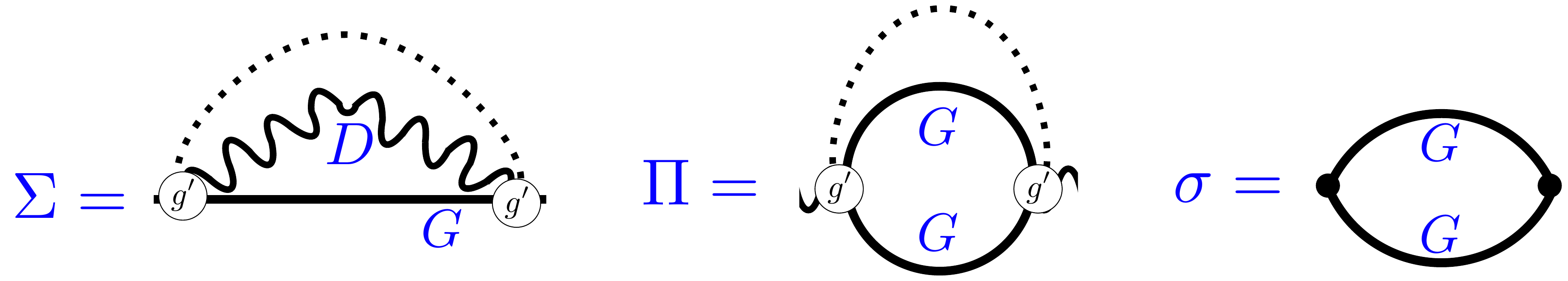}\\
    \caption{Feynman diagrams for the $\psi$ self energy $\Sigma$, and the $\phi$ self energy $\Pi$. The wavy line is the $\phi$ Green's function, and the smooth line is the $\psi$ Green's function. All Green's functions include self energy corrections. The dashed line represents an average over spatial disorder. All Green's functions and self energies become $2 \times 2$ matrices in the superconducting phase. }
    \label{fig:yukawa}
\end{figure}

After a disorder average, and in the limit of a large number of flavors where the saddle-point approximation is exact, we obtain the following SYK-type equations
for the electron Green's function $\hat{G}$ (a matrix in Nambu space) and the boson Green's function $D$:
\begin{subequations}
    \begin{eqnarray}
  \hat{G}(i\omega) &=& \int\frac{\rd^2 k}{(2\pi)^2}\left(i\omega\sigma_0+\mu\sigma_3-\varepsilon_k\sigma_3-\hat{\Sigma}(i\omega)\right)^{-1}, \label{eq:saddle_kspace1}\\
  D(i\nu) &=& \int\frac{\rd^2 q}{(2\pi)^2}\frac{1}{\nu^2+\omega_q^2-\Pi(i\nu)},\label{eq:saddle_kspace2} \\
  \Sigma(\tau) &=& g'^2G(\tau)D(\tau)+v^2G(\tau),\label{eq:saddle_kspace3} \\
  \Phi(\tau) &=& -(1-\alpha)(g'^2F(\tau)D(\tau)+v^2F(\tau)),\label{eq:saddle_kspace33} \\
  %\Pi(\tau) &=& -g'^2 \left({\rm tr}[\bar{G}(\tau)\bar{G}(-\tau)]-(1-\alpha){\rm tr}[\bar{F}(\tau)\bar{F}^+(-\tau)]\right),\label{eq:saddle_kspace4}\\
  \Pi(\tau) &=& -2g'^2 \left(G(\tau)G(-\tau)-(1-\alpha)F(\tau)F^\dagger(-\tau)\right),\label{eq:saddle_kspace4}\\
  \frac{1}{\gamma}&=&T\sum_{\nu} \int\frac{\rd^2 q}{(2\pi)^2}\frac{1}{\nu^2+\omega_q^2-\Pi(i\nu)}\,. \label{eq:saddle_constraint}
\end{eqnarray}
\label{eq:Yukawa_saddle}%
\end{subequations}
Here the $\sigma_i$ are Pauli matrices in Nambu space, and we use the following parametrization for $\hat{G}$ and $\hat{\Sigma}$
\begin{subequations}
    \begin{eqnarray}
        \hat{G}(\iw)&=&G(\iw)\sigma_0+F(\iw)\sigma_1,\\
        \hat{\Sigma}(\iw)&=&\Sigma(\iw)\sigma_0+\Phi(\iw)\sigma_1\,.
    \end{eqnarray}
\end{subequations}
These equations are indicated schematically in Figure~\ref{fig:yukawa} for $v=0$, including the diagrammatic representation of the current-current correlation function used to determine the conductivity $\sigma$.  Different from the translationally  invariant model,  the self-energies are momentum independent as a result of the extra $\delta$-function in (\ref{eq:distribution}) and (\ref{eq:vdistribution}). 

\begin{figure}[h]
    \centering
    \includegraphics[width=0.8\textwidth]{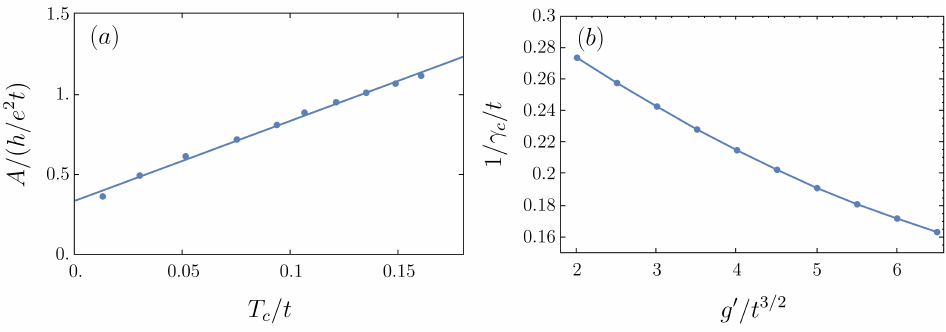}
    \caption{(a) Relationship between slope $A$ of the linear in $T$ resistivity and transition temperature $T_c$. Here $T_c$ is varied by varying $g'$ while keeping the system at the QCP. (b) The (inverse) value of the tuning parameter $\gamma$ at the critical point as a function of $g'$. }
    \label{fig:Tc}%
\end{figure}

~\\
{\it Appendix: Correlation between $T_c$ and the slope of linear in $T$ resistivity} -- We have also studied the correlation between $A$ -- the slope of the $T$ linear resistivity -- and $T_c$ when the system is tuned precisely to the QCP at $\gamma_c$ ($M=0$). In this case $T_c$ is tuned by varying the interaction strength $g'$. We find an essentially linear relationship, as may be seen in Fig.~\ref{fig:Tc}(a). The corresponding tuning parameter $\gamma_c$ as a function of $g'$ is shown in Fig. \ref{fig:Tc}(b).

%In Fig. \ref{fig:Tc}(a), we show the transition temperature $T_c$ as a function of $g'$ when the system is tuned to the QCP at $\gamma_c$ ($M=0$). Fig. \ref{fig:Tc}(b) shows $\gamma_c$ as a function of $g'$. We find $T_c$ is monotonically increasing with $g'$.

~\\
{\it Appendix: Superfluid stiffness} -- The superfluid stiffness may be obtained from the expectation value of the electronic kinetic energy $K$ and the Matsubara current-current correlation function $\Lambda(i\omega_n)$ (see (\ref{eq:SrhoK})) as \cite{Scalapino93}
    \be
    \frac{\rho_s}{\pi e^2} = \langle -K\rangle /2 - \Lambda(i\omega_n = 0). \label{app:rhoK}
    \ee
This result may be expressed purely in terms of the anomalous Green's function:
    \be
     \frac{\rho_s}{\pi e^2} = 2 T \sum_n \int \dd\epsilon \rho_\text{tr}(\epsilon) F^\dag(\epsilon, i\omega_n) F(\epsilon, i\omega_n).
    \ee
The detailed derivation of this result may be found in the supplement. 
\newpage
\foreach \x in {1,...,15}
{
\clearpage
\includepdf[pages={\x},angle=0]{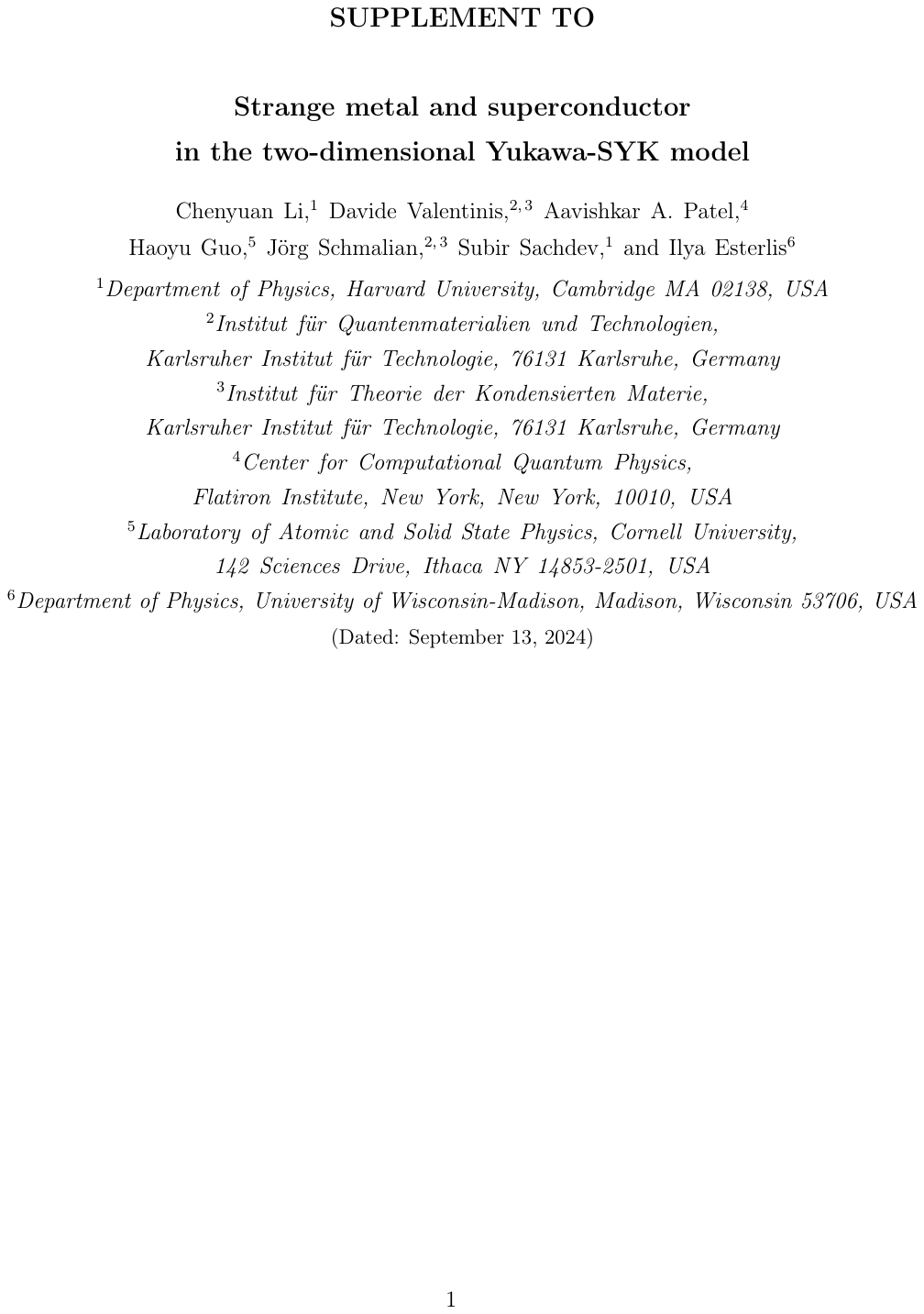} 
}

\end{document}